\pdfoutput=1
\documentclass[aps,prb,twocolumn,superscriptaddress,nofootinbib,floatfix,10pt]{revtex4-2}

\usepackage{amsmath,amssymb,amsfonts,bm}
\usepackage{booktabs}
\usepackage{graphicx}
\usepackage{xcolor}
\usepackage[colorlinks=true,linkcolor=blue!55!black,citecolor=blue!55!black,urlcolor=blue!55!black]{hyperref}

\newcommand{\dd}{\mathrm{d}}
\newcommand{\tr}{^{\dagger}}
\newcommand{\norm}[1]{\left\lVert#1\right\rVert}

\makeatletter
\g@addto@macro\acknowledgments{\interlinepenalty=10000}
\def\frontmatter@abstractheading{%
  \par\vspace{6pt}%
  {\centering\normalsize\bfseries Abstract\par}%
  \vspace{2pt}%
}
\makeatother

\begin{document}
\makeatletter
\let\orig@addtocontents\addtocontents
\renewcommand{\addtocontents}[2]{\ifnum\pdfstrcmp{#1}{toc}=0\else\orig@addtocontents{#1}{#2}\fi}
\makeatother

\title{High-accuracy reconstruction of machine-learning Hamiltonians in a fixed atomic-orbital representation from plane-wave projector augmented-wave calculations}

\author{Qinghong Wu}
\affiliation{School of Physics and Optoelectronic Engineering, Guangdong University of Technology, Guangzhou 510006, China}
\affiliation{School of Materials and Energy, Guangdong University of Technology, Guangzhou 510006, China}

\author{Jun Kang}
\affiliation{Beijing Computational Science Research Center, Beijing 100193, China}

\author{Minru Wen}
\email{mrwen@gdut.edu.cn}
\affiliation{School of Physics and Optoelectronic Engineering, Guangdong University of Technology, Guangzhou 510006, China}

\author{Huafeng Dong}
\affiliation{School of Physics and Optoelectronic Engineering, Guangdong University of Technology, Guangzhou 510006, China}

\author{Le Huang}
\email{huangle@gdut.edu.cn}
\affiliation{School of Physics and Optoelectronic Engineering, Guangdong University of Technology, Guangzhou 510006, China}
\affiliation{School of Materials and Energy, Guangdong University of Technology, Guangzhou 510006, China}

\author{Zhongming Wei}
\email{zmwei@semi.ac.cn}
\affiliation{State Key Laboratory of Semiconductor Physics and Chip Technologies, Institute of Semiconductors, Chinese Academy of Sciences, Beijing 100083, China}

\begin{abstract}
Machine-learning Hamiltonian (MLH) methods enable large-scale first-principles electronic-structure simulations, yet their equivariant models require sparse real-space matrices in a fixed localized atomic-orbital (AO) representation, creating a substantial gap with the widely used plane-wave projector augmented-wave (PW-PAW) formalism. Existing conversion schemes in compact fixed AO spaces suffer from metric mismatch, high-energy complement aliasing, and variational bias, and lack a unified treatment spanning magnetism, strong correlation, spin-orbit coupling, and hybrid functionals. Here we establish a unified framework for reconstructing accurate, sparse effective AO Hamiltonians directly from PW-PAW calculations. By decoupling the PAW projection metric from the purely geometric two-center AO overlap and combining spectral-filtered direct operator projection, window-protected smooth complement completion, and energy-origin-covariant Fourier interpolation, our approach preserves strict orbital-gauge consistency while faithfully recovering low-energy spectra and wavefunctions. Systematic VASP benchmarks across moir\'e superlattices, ferromagnetic metals, strongly correlated oxides, noncollinear frustrated magnets, and hybrid functionals generally achieve millielectronvolt-scale off-grid band accuracy while substantially reducing wavefunction action residuals. End-to-end equivariant graph neural network training on semilocal and hybrid-functional perturbed structures further confirms smooth cross-configuration learnability, converging to the direct reconstruction limit on unseen geometries from atomic coordinates alone and bridging PW-PAW calculations with cross-functional MLH modeling.
\end{abstract}

\maketitle

\section{Introduction}\label{sec:intro}

In recent years, machine-learning Hamiltonian (MLH) methods, represented by DeepH and HamGNN, have become an important research direction in electronic-structure calculations~\cite{li2022deeph,gong2023deephe3,zhong2023hamgnn}. These methods typically rely on $E(3)$-equivariant graph neural networks to directly learn the mapping between local atomic configurations and Hamiltonian matrix elements $H_{\mu\nu}(\mathbf{R})$ in a localized numerical atomic orbital (NAO) or pseudo-atomic orbital (PAO) representation. Owing to the real-space locality of atomic-orbital bases~\cite{junquera2001,ozaki2003,ozaki2004}, the corresponding Hamiltonian $H_{\mu\nu}(\mathbf{R})$ and overlap $S_{\mu\nu}(\mathbf{R})$ matrices exhibit a compactly supported sparse structure, enabling MLH models to predict and construct the electronic structure of large-scale systems with nearly $O(N_a)$ scaling. This property provides a new computational route for electronic-structure studies of complex systems such as moir\'e superlattices, supercell defect systems, and amorphous materials, as well as excited-state nonadiabatic couplings and dynamics simulations~\cite{zhong2023hamgnn,xu2025prx,dai2026npj,zhang2026prl}.

At present, MLH data generation predominantly relies on first-principles codes that natively support localized atomic-orbital bases, such as OpenMX~\cite{ozaki2003,ozaki2004}, ABACUS~\cite{li2016abacus}, and SIESTA~\cite{soler2002siesta}. However, the plane-wave projector augmented-wave (PAW) method, represented by VASP~\cite{kresse1996}, offers systematically controllable plane-wave kinetic-energy cutoffs, broad support for exchange-correlation functionals, and mature density-functional perturbation theory (DFPT) algorithms~\cite{blochl1994,kresse1999,baroni2001dfpt,gajdos2006paw}, and has accumulated a large body of high-quality electronic-structure data for materials. Accurately converting plane-wave PAW calculation results into atomic-orbital Hamiltonians with a fixed orbital gauge and compact real-space support is therefore a key challenge for expanding the training-data sources and application scope of MLH methods.

Existing methods for converting plane-wave representations into localized atomic-orbital representations still struggle to simultaneously meet the MLH requirements of orbital-gauge consistency, low-energy spectral accuracy, and operator completeness. Variational disentanglement approaches based on maximally localized Wannier functions (MLWFs) achieve high-accuracy band interpolation~\cite{marzari2012}, but their orbital shapes, local orientations, and gauges generally depend on the specific atomic configuration, making it difficult to construct an atomic-orbital representation with a fixed spherical-harmonic tensor structure for unified training and cross-configuration transfer of $E(3)$-equivariant neural networks. Finite-band projection methods onto fixed trial atomic orbitals~\cite{sanchez1995,sanchez1996}, such as LOBSTER~\cite{maintz2013,maintz2016,nelson2020} and PAOFLOW~\cite{agapito2013,agapito2016,nardelli2018}, preserve consistent orbital orientations and angular-momentum channels, yet face three challenges: (1)~\emph{Dual-metric mismatch}: the PAW projection metric $S_P(\mathbf{k})$ differs from the target atomic-orbital overlap metric $S_A(\mathbf{R})$; (2)~\emph{High-energy complement aliasing}: nonunitary projection and constant-level approximations for the orthogonal complement may introduce unphysical level repulsion and off-grid dispersion ringing; and (3)~\emph{Incomplete operator information}: fitting eigenvalues alone cannot determine Hamiltonian couplings within and outside the target subspace, leaving the reconstructed operator insufficiently constrained.

The recently proposed real-space integration method HPRO~\cite{gong2024hpro} enables direct reconstruction of localized atomic-orbital Hamiltonians from plane-wave calculations without explicitly summing over large numbers of high-energy unoccupied bands, and discusses its extension to the PAW formalism. However, when applied to the compact fixed atomic-orbital bases required by MLH models, the finite orbital space can still introduce variational basis errors, causing systematic shifts in the low-energy Ritz eigenvalues relative to the plane-wave reference. Furthermore, a systematic framework for the unified extension of such methods to PAW collinear magnetism, DFT+$U$~\cite{dudarev1998,liechtenstein1995,bengone2000}, noncollinear magnetism~\cite{hobbs2000ncl}, spin-orbit coupling (SOC), and screened hybrid functionals~\cite{heyd2003,krukau2006,paier2006} remains lacking.

To address these issues, we propose a unified effective-Hamiltonian reconstruction framework mapping plane-wave PAW calculations onto a fixed atomic-orbital representation for MLH construction, aimed at obtaining sparse effective Hamiltonians $H_A(\mathbf{R})$ that combine physical accuracy, orbital-gauge consistency, and real-space locality. First, by distinguishing the PAW projection metric $S_P(\mathbf{k})$ from the target atomic-orbital overlap metric $S_A(\mathbf{R})$, we establish a dual-metric description and combine direct PAW operator assembly with spectral-filtered Ritz regularization incorporating $\tanh$ high-energy spectral clipping to improve low-energy spectral accuracy in compact orbital spaces. Second, for noncollinear spinor and hybrid-functional systems, we propose a window-protected smooth complement-completion strategy and a semilocal-anchored dual-mesh interpolation scheme, respectively, to enhance the applicability of the reconstruction framework across different magnetic states and exchange-correlation functionals. In addition, we formulate an energy-origin-covariant Fourier transform and derive a three-term off-grid error attribution identity to provide a quantitative basis for analyzing interpolation accuracy and individual error sources.

We carry out systematic numerical benchmarks on PW-PAW VASP calculations across diverse material classes and exchange-correlation functionals, encompassing same-source comparisons with external codes, off-grid spectral and operator-action diagnostics, and end-to-end cross-configuration machine-learning and transfer tests to evaluate the accuracy, stability, and generalization capability of the proposed framework. The remainder of this paper is organized as follows. Section~\ref{sec:method} introduces the theoretical framework and numerical methodology. Section~\ref{sec:results} presents the systematic numerical validation results and cross-configuration transfer tests. Section~\ref{sec:conclusion} summarizes and discusses the main findings.

\section{Theory and methodology}\label{sec:method}

\subsection{Atomic-orbital basis and dual metrics}\label{sec:basis_metrics}

Let $\mathbf{a}$ denote the lattice-vector matrix with unit-cell volume $\Omega=|\det\mathbf{a}|$, $\mathbf{R}\in\mathbb{Z}^3$ a lattice translation vector, $\bm{\tau}_a$ the Cartesian position of atom $a$ within the unit cell, and $\mathbf{k}$ a fractional reciprocal-space wave vector. Using the composite orbital index $\mu=(a,\zeta,l,m)$, a localized atomic orbital centered at $\bm{\tau}_a+\mathbf{R}\mathbf{a}$ is defined as
\begin{equation}
\phi_{\mu\mathbf{R}}(\mathbf{r}) = u_{\zeta l}(|\mathbf{r}-\bm{\tau}_a-\mathbf{R}\mathbf{a}|)\,Y^r_{lm}(\widehat{\mathbf{r}-\bm{\tau}_a-\mathbf{R}\mathbf{a}}),
\label{eq:ao_def}
\end{equation}
where $\zeta$ labels radial multiplicities within angular-momentum channel $l$, $Y^r_{lm}$ are real spherical harmonics ordered by $m=-l,\ldots,+l$, and the radial functions are normalized within cutoff radius $r_c$ according to $\int_0^{r_c} r^2 |u_{\zeta l}(r)|^2\,\dd r = 1$. The total number of spatial orbitals per unit cell is $N_{\mathrm{AO}}$. We adopt the cell-periodic Fourier-transform convention $M_{\mu\nu}(\mathbf{k}) = \sum_{\mathbf{R}\in\mathbb{Z}^3} e^{2\pi i \mathbf{k}\cdot\mathbf{R}} M_{\mu\nu}(\mathbf{R})$. Below the plane-wave kinetic-energy cutoff $E_{\mathrm{cut}}$, with momentum transfer $\mathbf{q}=\mathbf{k}_{\mathrm{cart}}+\mathbf{G}$, the plane-wave expansion matrix $\Phi_{\mathbf{G}\mu}(\mathbf{k})$ of the trial atomic orbitals is evaluated analytically via the spherical Bessel transform,
\begin{equation}
\Phi_{\mathbf{G}\mu}(\mathbf{k}) = \frac{4\pi(-i)^l}{\sqrt{\Omega}} e^{-i\mathbf{q}\cdot\bm{\tau}_a} Y^r_{lm}(\hat{\mathbf{q}})\!\int_0^{r_c}\!\! r^2 u_{\zeta l}(r) j_l(|\mathbf{q}|r)\,\dd r.
\label{eq:pw_column}
\end{equation}

In Bl\"ochl's PAW formalism, the linear transformation $\mathcal{T}$ maps smooth pseudo-wavefunctions onto all-electron states, yielding the generalized Kohn--Sham equation $\widetilde{H}(\mathbf{k})C(\mathbf{k}) = O(\mathbf{k})C(\mathbf{k})E(\mathbf{k})$ with $\widetilde{H}(\mathbf{k}) = \mathcal{T}\tr H^{\mathrm{AE}}\mathcal{T}$ and overlap operator $O(\mathbf{k}) = \mathcal{T}\tr\mathcal{T} = I + P(\mathbf{k})\Delta P\tr(\mathbf{k}) \succ 0$, where $C(\mathbf{k})$ satisfies $C\tr(\mathbf{k})O(\mathbf{k})C(\mathbf{k})\simeq I$, $E(\mathbf{k})=\mathrm{diag}(E_{1\mathbf{k}},\ldots,E_{N_b\mathbf{k}})$, $P(\mathbf{k})$ collects the nonlocal projectors, and $\Delta$ is the one-center augmentation overlap tensor. Projecting onto the $N_{\mathrm{AO}}$ trial orbitals $\Phi(\mathbf{k})$ defines the positive-definite \emph{PAW projection metric} $S_P(\mathbf{k})$ and band-projection matrix $A(\mathbf{k})$,
\begin{equation}
S_P(\mathbf{k}) = \Phi\tr(\mathbf{k}) O(\mathbf{k})\Phi(\mathbf{k}),\qquad
A(\mathbf{k}) = \Phi\tr(\mathbf{k}) O(\mathbf{k}) C(\mathbf{k}).
\label{eq:paw_metric}
\end{equation}
Because $S_P(\mathbf{k})$ depends on the plane-wave cutoff and PAW augmentation, it is used exclusively during wavefunction and operator projection. To decouple the target representation from the plane-wave basis, we introduce the purely geometric \emph{real-space two-center overlap metric} $S_A(\mathbf{R})$ on the same $N_{\mathrm{AO}}$ orbitals,
\begin{equation}
S_{A,\mu\nu}(\mathbf{R}) = \int_{\mathbb{R}^3} \phi_{\mu\mathbf{0}}^*(\mathbf{r})\,\phi_{\nu\mathbf{R}}(\mathbf{r})\,\dd^3\mathbf{r},
\label{eq:sa_def}
\end{equation}
evaluated via bond-axis two-dimensional quadrature and real spherical-harmonic rotation. Transforming between the PAW metric $S_P(\mathbf{k})$ and geometric metric $S_A(\mathbf{k})$ proceeds through a common L\"owdin-orthonormalized effective operator $h(\mathbf{k})$~\cite{lowdin1950} via the isospectral, $O(3)$-equivariant congruence $M(\mathbf{k})=S_A(\mathbf{k})^{1/2}S_P(\mathbf{k})^{-1/2}$ ($M S_P M\tr = S_A$; see Sec.~\ref{sec:supp_projection} of the Supplemental Material~\cite{SM} for full derivations and proofs). During MLH inference on unseen structures, $S_A(\mathbf{R})$ follows directly from atomic coordinates, so the network only predicts $H_A(\mathbf{R})$.

To eliminate dependence on external basis libraries while overcoming scattering tails and short cutoff radii in native VASP PAW pseudo partial waves, we construct compactly supported numerical atomic orbitals directly from the native PAW reference partial waves via smooth tapering, Slater polarization augmentation, and intra-channel Gram--Schmidt orthonormalization, with diffuse Gaussian mixing for vacuum-extended two-dimensional conduction states (see Sec.~\ref{sec:supp_basis} of the Supplemental Material).

\subsection{Direct operator projection and spectral filtering}\label{sec:direct_ritz}

In conventional finite-band spectral summation, $H_{\mathrm{spec}}(\mathbf{k}) = A(\mathbf{k})[E(\mathbf{k})-\eta I]A\tr(\mathbf{k}) + \eta S_P(\mathbf{k})$, low-projectability high-energy conduction states induce perturbative level repulsion, while replacing the orthogonal complement by a constant shift $\eta$ produces dispersion ringing at off-grid $\mathbf{k}$ points. To capture the true operator coupling of the infinite orthogonal complement without explicitly diagonalizing large numbers of unoccupied bands~\cite{qian2008}, we project the plane-wave PAW Hamiltonian directly onto the trial AO subspace. For nonmagnetic semilocal, collinear spin-polarized, and DFT+$U$ calculations with local correlation corrections, the self-consistent one-center Hubbard potential corrections are already absorbed into the spin-resolved PAW strength matrix $D_\sigma$. Letting $\sigma\in\{\uparrow,\downarrow\}$ with $s_\uparrow=+1$ and $s_\downarrow=-1$, the action of the generalized PAW Hamiltonian $\widetilde{H}_\sigma(\mathbf{k})$ on $\Phi(\mathbf{k})$ decomposes into kinetic, soft-grid local-potential fast Fourier transform, and separable nonlocal terms,
\begin{align}
\widetilde{H}_\sigma(\mathbf{k})\Phi &= T(\mathbf{k})\Phi + P(\mathbf{k})D_\sigma P\tr(\mathbf{k})\Phi \nonumber\\
&\quad + \Pi_{\mathbf{G}}\mathcal{F}\Bigl[\bigl(V_0(\mathbf{r})+s_\sigma B_z(\mathbf{r})\bigr)\mathcal{F}^{-1}\bigl(\Pi_{\mathbf{G}}\tr\Phi\bigr)\Bigr], \label{eq:direct_paw_action}\\
H_{P,\sigma}(\mathbf{k}) &= \frac{1}{2}\Bigl[\Phi\tr\widetilde{H}_\sigma(\mathbf{k})\Phi + \bigl(\Phi\tr\widetilde{H}_\sigma(\mathbf{k})\Phi\bigr)\tr\Bigr], \label{eq:hp_def}
\end{align}
where $\Pi_{\mathbf{G}}$ embeds the wavefunction sphere into the real-space soft FFT grid, and $V_0(\mathbf{r})$ and $B_z(\mathbf{r})$ denote the scalar local potential and collinear exchange field, respectively.

Because a compact fixed AO basis $\Xi(\mathbf{k})=\Phi(\mathbf{k}) S_P(\mathbf{k})^{-1/2}$ does not span the full plane-wave space, low-energy Ritz eigenvalues of the directly orthonormalized operator $K_{0,\sigma}(\mathbf{k})=S_P(\mathbf{k})^{-1/2}H_{P,\sigma}(\mathbf{k})S_P(\mathbf{k})^{-1/2}$ exhibit variational overestimation due to high-energy admixture. To remove this bias, motivated by smooth-window projection strategies~\cite{ozaki2024}, we introduce a spectral-filtered trial subspace $F(\hat{h}_\sigma)\Xi$ with a strictly positive regularization floor $0<\beta\le 1$ and smooth energy weight
\begin{equation}
\begin{aligned}
x_n &= \mathrm{clip}\!\left(\frac{E_t - E_{n\mathbf{k}\sigma}}{\Delta_d},\,0,\,1\right),\quad
F(E_{n\mathbf{k}\sigma})^2 = \beta + w_{n\sigma},\\
w_{n\sigma} &= \frac{(1-\beta)x_n^2(3-2x_n)}{1+\exp[(E_{n\mathbf{k}\sigma}-E_c)/\Delta_b]},
\end{aligned}
\label{eq:filter_weight}
\end{equation}
where $E_t=\min_{\mathbf{k}} E_{N_b\mathbf{k}\sigma}$ is the lower bound of the highest computed band, $E_c$ is the filter center above the target energy window, and $\Delta_b,\Delta_d$ are smooth transition widths. With $V_\sigma(\mathbf{k})=S_P(\mathbf{k})^{-1/2}A_\sigma(\mathbf{k})$, completeness rigorously folds the entire uncomputed infinite orthogonal complement into the $\beta$ floor of $K_{0,\sigma}$, yielding the filtered overlap moment $G_\sigma(\mathbf{k})$, Hamiltonian moment $K_\sigma(\mathbf{k})$, and L\"owdin-orthonormalized effective operator $h_\sigma(\mathbf{k})$:
\begin{align}
G_\sigma(\mathbf{k}) &= \beta I + V_\sigma(\mathbf{k})\,\mathrm{diag}(w_{n\sigma})\,V_\sigma\tr(\mathbf{k}), \label{eq:filtered_G}\\
K_\sigma(\mathbf{k}) &= \beta K_{0,\sigma}(\mathbf{k}) + V_\sigma(\mathbf{k})\,\mathrm{diag}(w_{n\sigma}E_{n\mathbf{k}\sigma})\,V_\sigma\tr(\mathbf{k}), \label{eq:filtered_K}\\
h_\sigma(\mathbf{k}) &= G_\sigma(\mathbf{k})^{-1/2}\,K_\sigma(\mathbf{k})\,G_\sigma(\mathbf{k})^{-1/2}. \label{eq:filtered_h}
\end{align}
Because $\beta I\preceq G_\sigma(\mathbf{k})\preceq I$, the filtered subspace retains full rank $N_{\mathrm{AO}}$ with well-conditioned orthonormalization and unitary invariance within degenerate subspaces.

In itinerant metals and strongly correlated DFT+$U$ oxides, unresolved high-energy Ritz eigenvalues $\lambda_i\gg E_F$ of $h_\sigma(\mathbf{k})=X_\sigma(\mathbf{k})\Lambda_\sigma(\mathbf{k})X_\sigma\tr(\mathbf{k})$ can reach tens of electronvolts and amplify small angular interpolation errors during Fourier interpolation. In the spirit of spectral transformations prior to band interpolation~\cite{wu2025ht}, we suppress this amplification via a $C^2$-continuous monotonic spectral clipping map above threshold $E_{\mathrm{clip}}$ with width $\Delta_{\mathrm{clip}}$,
\begin{equation}
f_{\mathrm{clip}}(x) = \begin{cases}
x, & x \le E_{\mathrm{clip}}, \\
E_{\mathrm{clip}} + \Delta_{\mathrm{clip}} \tanh\!\left(\dfrac{x-E_{\mathrm{clip}}}{\Delta_{\mathrm{clip}}}\right), & x > E_{\mathrm{clip}},
\end{cases}
\label{eq:tanh_clip}
\end{equation}
which leaves the full orthonormal eigenbasis $X_\sigma(\mathbf{k})$ and every eigenpair below $E_{\mathrm{clip}}$ strictly unchanged while bounding the high-energy spectrum. Pulling back through $S_A(\mathbf{k})^{1/2}$ gives the target AO Hamiltonian,
\begin{equation}
H_{A,\sigma}(\mathbf{k}) = S_A(\mathbf{k})^{1/2}\,X_\sigma(\mathbf{k})\,f_{\mathrm{clip}}\!\bigl(\Lambda_\sigma(\mathbf{k})\bigr)\,X_\sigma\tr(\mathbf{k})\,S_A(\mathbf{k})^{1/2}.
\label{eq:ha_pullback}
\end{equation}
All threshold-selection criteria, proofs of full-rank target-subspace preservation, and the complete reconstruction procedure are detailed in Sec.~\ref{sec:supp_filter_proof} of the Supplemental Material.

\subsection{Smooth complement completion and dual-mesh interpolation}\label{sec:locality_hse}

For screened hybrid functionals containing nonlocal exact exchange as well as two-component noncollinear spinor and SOC calculations, the self-consistent eigenpairs $\bigl(E(\mathbf{k}), C(\mathbf{k})\bigr)$ already incorporate nonlocal exact-exchange and relativistic spin-coupling effects. At each source wave vector, we retain $m_{\mathbf{k}}$ bands spanning the frozen target window~\cite{souza2001} plus a small high-energy buffer (completing boundary degenerate multiplets), partition the normalized projection columns into target-window columns $V_T$ and buffer columns $V_U$, and construct an orthonormal frame $Q(\mathbf{k})\in\mathbb{C}^{N_{\mathrm{AO}}\times m_{\mathbf{k}}}$ via \emph{window-first two-step polar orthonormalization},
\begin{equation}
Q_T = \mathrm{polar}(V_T),\quad
Q_U = \mathrm{polar}\bigl[(I - Q_T Q_T\tr)V_U\bigr],
\label{eq:protected_polar}
\end{equation}
reassembled in band order. This hierarchical orthogonalization strictly protects the target-window subspace $Q_T$ from rotations induced by high-energy buffer states. Letting $Z(\mathbf{k})$ denote an orthonormal basis for the orthogonal complement of $Q(\mathbf{k})$, the Hermitian orthonormal-frame effective Hamiltonian $h(\mathbf{k})$~\cite{kvaal2008} is constrained to the convex feasible set
\begin{equation}
\begin{aligned}
h(\mathbf{k}) &= h\tr(\mathbf{k}),\quad
h(\mathbf{k})Q(\mathbf{k}) = Q(\mathbf{k})E_{1:m_{\mathbf{k}}}(\mathbf{k}),\\
Z\tr(\mathbf{k}) h(\mathbf{k}) Z(\mathbf{k}) &\succeq \bigl(E_{m_{\mathbf{k}}\mathbf{k}} + \Delta_c\bigr)I,
\end{aligned}
\label{eq:feasible_set}
\end{equation}
with complement gap margin $\Delta_c>0$ preventing complement states from intruding into the target window as spurious bands. Exploiting the duality between real-space locality and reciprocal-space smoothness, we minimize a strictly convex Voronoi-area-weighted reciprocal-space smoothness functional over Eq.~\eqref{eq:feasible_set} via projected gradient descent:
\begin{equation}
\begin{aligned}
\mathcal{J}_{\mathrm{loc}}[\{h(\mathbf{k})\}] &= \frac{1-\gamma}{4}\sum_{\mathbf{k},\mathbf{b}} w_{\mathbf{b}} \norm{h(\mathbf{k}) - \Theta_{\mathbf{b}}\tr h(\mathbf{k}+\mathbf{b}) \Theta_{\mathbf{b}}}_F^2 \\
&\quad + \frac{\gamma}{2}\sum_{\mathbf{k}}\norm{h(\mathbf{k}) - \eta_0 I}_F^2,
\end{aligned}
\label{eq:voronoi_objective}
\end{equation}
where $(\Theta_{\mathbf{b}})_{\mu\mu} = \exp(2\pi i\,\mathbf{b}\cdot\mathbf{t}_\mu)$ enforces gauge covariance with respect to fractional orbital centers $\mathbf{t}_\mu$, $w_{\mathbf{b}}$ are normalized Voronoi facet weights, and $\gamma>0$, $\eta_0$ provide mild regularization toward a high-energy reference. The unique global minimizer is independent of the choice of complement basis $Z(\mathbf{k})$ (full convexity, uniqueness, and basis-invariance proofs, parameter choices, and the numerical optimization procedure are given in Sec.~\ref{sec:supp_voronoi_proof} of the Supplemental Material). For noncollinear spinor and SOC calculations, the trial space expands to $2N_{\mathrm{AO}}$ spinor orbitals with geometric overlap $S_s(\mathbf{R})=I_2\otimes S_A(\mathbf{R})$; after aligning the spin quantization axis to the global Cartesian frame via an $SU(2)$ spinor rotation (Sec.~\ref{sec:supp_spinor} of the Supplemental Material), Eq.~\eqref{eq:voronoi_objective} is solved globally in spinor space, preserving off-diagonal spin-flip blocks $H_A^{\uparrow\downarrow}(\mathbf{R})$.

For screened hybrid functionals computed on coarse source $\mathbf{k}$ meshes, a constant prior $\eta_0 I$ alone cannot resolve high-frequency reciprocal-space components originating from longer-range semilocal hoppings. Analogous to interpolating nonlocal self-energy corrections relative to a semilocal baseline in a localized basis~\cite{kotani2007}, we introduce a \emph{fine-grid semilocal-anchored dual-mesh reconstruction}. First, an anchor operator $H_A^{\mathrm{anc}}(\mathbf{R}) = H_A^{\mathrm{PBE}}(\mathbf{R}) + (E_F^{\mathrm{HSE}} - E_F^{\mathrm{PBE}}) S_A(\mathbf{R})$ and its Fourier sum $H_A^{\mathrm{anc}}(\mathbf{k})$ are built from a fine-grid semilocal PBE calculation using identical geometry and AO basis. At each coarse HSE06 grid point $\mathbf{k}\in K_{\mathrm{HSE}}$, we form $h^{\mathrm{anc}}(\mathbf{k}) = S_A(\mathbf{k})^{-1/2} H_A^{\mathrm{anc}}(\mathbf{k}) S_A(\mathbf{k})^{-1/2}$ and minimize the reciprocal-space smoothness functional with respect to the nonlocal exchange-correlation difference $\Delta h(\mathbf{k}) = h(\mathbf{k}) - h^{\mathrm{anc}}(\mathbf{k})$ subject to the exact HSE06 eigenpair constraints~\eqref{eq:feasible_set}. The smooth correction is then transformed back to real space and superposed onto the fine-grid anchor:
\begin{equation}
\begin{aligned}
H_A(\mathbf{R}) &= H_A^{\mathrm{anc}}(\mathbf{R}) \\
&\quad + \mathcal{F}_{K_{\mathrm{HSE}}\to\mathbf{R}}\!\Bigl[S_A(\mathbf{k})^{1/2}\Delta h(\mathbf{k})S_A(\mathbf{k})^{1/2}\Bigr](\mathbf{R}).
\end{aligned}
\label{eq:hse_final_hr}
\end{equation}

\subsection{Fourier interpolation and off-grid error attribution}\label{sec:realspace_attribution}

\begin{figure*}[t]
\centering
\includegraphics[width=\textwidth]{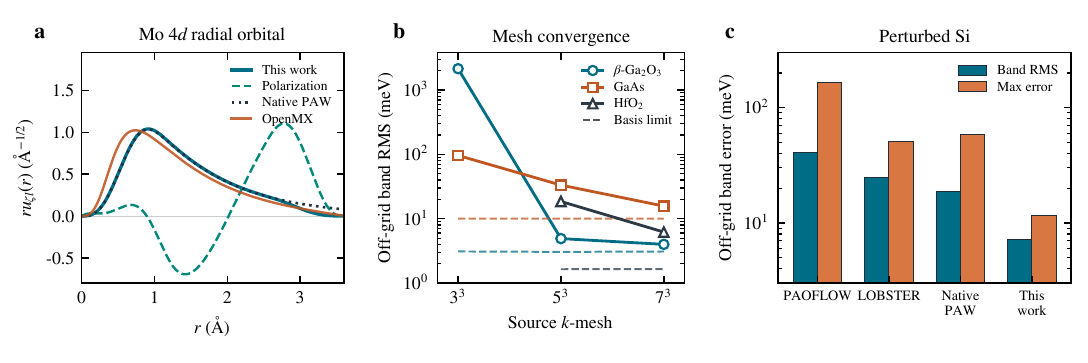}
\caption{(a) Radial profiles of the $\mathrm{Mo}$ $4d$ valence orbital and polarization shell from this work compared with the native VASP PAW partial wave and OpenMX orbital. (b) Off-grid band RMS error (solid) versus source $\mathbf{k}$-mesh density relative to the finite-basis limit (dashed). (c) Off-grid band errors on perturbed $\mathrm{Si}$ across reconstruction schemes.}
\label{fig:basis_attribution}
\end{figure*}

Under a rigid shift of the energy origin, $E_{n\mathbf{k}}\mapsto E_{n\mathbf{k}}+c$, the reciprocal-space Hamiltonian transforms as $H_A(\mathbf{k})\mapsto H_A(\mathbf{k})+c S_A(\mathbf{k})$. If $H_A(\mathbf{k})$ is directly inverse Fourier transformed on a finite source mesh and subsequently diagonalized alongside the exact geometric overlap $S_A(\mathbf{k}')$ at an off-grid point $\mathbf{k}'\notin K_{\mathrm{src}}$, any energy-origin shift introduces an aliasing error proportional to $c[\widetilde{S}_A(\mathbf{k}')-S_A(\mathbf{k}')]$, where $\widetilde{S}_A(\mathbf{k}')$ denotes the undersampled Fourier-interpolated overlap. To guarantee exact \emph{energy-origin shift covariance}, we introduce a reference energy $\eta=E_F$ tied to the source calculation and Fourier transform only the offset operator $\mathcal{D}(\mathbf{k})=H_A(\mathbf{k})-\eta S_A(\mathbf{k})$, distributing each modulo-grid translation class $\mathbf{t}$ evenly across its $g_{ab}(\mathbf{t})$ shortest Wigner-Seitz images measured by bond length $|\bm{\tau}_b+\mathbf{R}\mathbf{a}-\bm{\tau}_a|$~\cite{pizzi2020}:
\begin{align}
B_{\mu\nu}(\mathbf{t}) &= \frac{1}{N_k}\sum_{\mathbf{k}\in K_{\mathrm{src}}} e^{-2\pi i \mathbf{k}\cdot\mathbf{t}}\,\bigl[H_{A,\mu\nu}(\mathbf{k}) - \eta S_{A,\mu\nu}(\mathbf{k})\bigr], \label{eq:offset_dft}\\
H_{A,\mu\nu}(\mathbf{R}) &= \eta S_{A,\mu\nu}(\mathbf{R}) + \frac{e^{-2\pi i \mathbf{k}_0\cdot(\mathbf{R}-\mathbf{t})}}{g_{ab}(\mathbf{t})}\,B_{\mu\nu}(\mathbf{t}), \label{eq:gauge_covariant_hr}
\end{align}
where $\mathbf{k}_0$ is the fractional grid shift and the first term uses the exact two-center overlap across the full support of $S_A(\mathbf{R})$. Under any rigid shift $c$, $\mathcal{D}(\mathbf{k})$ is strictly invariant and all generalized eigenvalues shift by exactly $c$ everywhere in the Brillouin zone.

To dissect error sources quantitatively at independent off-grid wave vectors $\mathbf{k}\notin K_{\mathrm{src}}$ with plane-wave reference eigenvalues $E_{n\mathbf{k}}$ and wavefunctions $C(\mathbf{k})$, we construct the same-$\mathbf{k}$ effective operator $h_\star(\mathbf{k})$ directly from $C(\mathbf{k})$ using identical filter parameters, and denote the orthonormal-frame matrices of the untruncated dense and final sparse models by $h_d(\mathbf{k})=S_A(\mathbf{k})^{-1/2}H_{A,\mathrm{dense}}(\mathbf{k})S_A(\mathbf{k})^{-1/2}$ and $h_s(\mathbf{k})=S_A(\mathbf{k})^{-1/2}H_{A,\mathrm{sparse}}(\mathbf{k})S_A(\mathbf{k})^{-1/2}$, respectively. For each ordered band $n$, the total off-grid error obeys an exact three-term scalar attribution identity:
\begin{equation}
\begin{aligned}
e_n\bigl(h_s(\mathbf{k})\bigr) - E_{n\mathbf{k}} &= \Bigl[e_n\bigl(h_\star(\mathbf{k})\bigr) - E_{n\mathbf{k}}\Bigr] \\
&\quad + \Bigl[e_n\bigl(h_d(\mathbf{k})\bigr) - e_n\bigl(h_\star(\mathbf{k})\bigr)\Bigr] \\
&\quad + \Bigl[e_n\bigl(h_s(\mathbf{k})\bigr) - e_n\bigl(h_d(\mathbf{k})\bigr)\Bigr],
\end{aligned}
\label{eq:three_term_eigen}
\end{equation}
whose right-hand terms isolate, respectively, the \emph{same-$\mathbf{k}$ finite-basis reconstruction error}, the \emph{reciprocal-space interpolation error}, and the \emph{real-space sparsity truncation error}. Complementarily, defining the isometric polar factor $Q(\mathbf{k})=\mathrm{polar}\bigl(V_e(\mathbf{k})\bigr)$ of the target-window PAW projection $V_e(\mathbf{k})=S_P(\mathbf{k})^{-1/2}A_e(\mathbf{k})$, we evaluate the \emph{wavefunction operator-action residual} and its analogous decomposition:
\begin{equation}
\begin{aligned}
L(\mathbf{k}) &\equiv h_s(\mathbf{k})Q(\mathbf{k}) - Q(\mathbf{k})E_e(\mathbf{k}) \\
&= \bigl[h_\star Q - Q E_e\bigr] + \bigl(h_d - h_\star\bigr)Q + \bigl(h_s - h_d\bigr)Q,
\end{aligned}
\label{eq:action_residual}
\end{equation}
simultaneously quantifying fidelity in both the eigenvalue spectrum and the eigenstate subspace orientation.

\section{Results and discussion}\label{sec:results}

To systematically assess the accuracy of our framework across diverse bonding environments, spin degrees of freedom, and exchange-correlation functionals, all benchmarks in this work are carried out on PW-PAW VASP calculations, ensuring strict consistency across reference wavefunctions, local potentials, and PAW one-center augmentation operators. All off-grid errors are measured directly against VASP reference bands on independent off-grid $\mathbf{k}$ points strictly excluding the source mesh and its symmetry equivalents, using the native VASP electrostatic-potential energy zero.

\subsection{Basis validation and method comparisons}\label{sec:res_attribution}

\begin{table}[t]
\centering
\caption{Off-grid band-edge mean absolute error and maximum error $\mathrm{MAE} / \Delta_{\max}$ (in meV) for three localized AO basis sets on same-source VASP PW-PAW calculations.}
\label{tab:ao_comparison}
\resizebox{\columnwidth}{!}{%
\begin{tabular}{lccc}
\toprule
System & Native VASP PAW & OpenMX basis & This work \\
\midrule
Monolayer $\mathrm{MoS_2}$ & $4.27\ /\ 17.28$ & $6.83\ /\ 20.57$ & $2.20\ /\ 9.77$ \\
Tetragonal $\mathrm{HfO_2}$ & $9.41\ /\ 57.37$ & $12.41\ /\ 50.08$ & $3.47\ /\ 27.58$ \\
Monoclinic $\mathrm{HfO_2}$ & $91.56\ /\ 334.15$ & $19.03\ /\ 64.54$ & $3.57\ /\ 19.61$ \\
Orthorhombic $\mathrm{CsPbI_3}$ & $9.64\ /\ 41.25$ & $45.75\ /\ 194.21$ & $4.42\ /\ 16.27$ \\
Twisted bilayer $\mathrm{MoS_2}$ & $8.54\ /\ 35.47$ & $20.50\ /\ 66.77$ & $7.56\ /\ 30.22$ \\
\bottomrule
\end{tabular}}
\end{table}

\begin{figure}[t]
\centering
\includegraphics[width=\columnwidth]{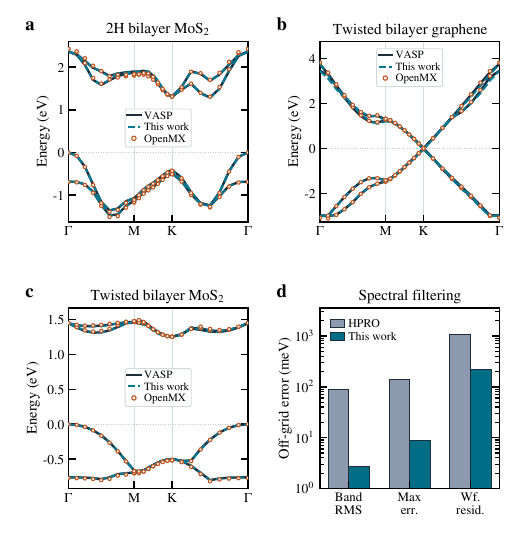}
\caption{Band structures of (a) 2H bilayer $\mathrm{MoS_2}$, (b) 28-atom twisted bilayer graphene, and (c) 42-atom twisted bilayer $\mathrm{MoS_2}$ from VASP, this work, and OpenMX. (d) Off-grid band errors and wavefunction residuals on 2H bilayer $\mathrm{MoS_2}$ for unfiltered HPRO and this work.}
\label{fig:bilayer_bands}
\end{figure}

Figure~\ref{fig:basis_attribution}(a) and Table~\ref{tab:ao_comparison} compare native VASP PAW pseudo partial waves, external OpenMX radial orbitals, and our compact numerical atomic orbitals on identical VASP pseudo-wavefunctions. Although the OpenMX basis shares the exact same orbital dimension as our basis, its radial functions are optimized for norm-conserving pseudopotentials, creating mismatches in near-core nodal positions and asymptotic-tail behavior with VASP PAW pseudo-wavefunctions that inflate off-grid band errors by a factor of $3\text{--}10$. Conversely, using raw native VASP PAW partial waves without polarization shells degrades accuracy in low-symmetry wide-gap oxides such as monoclinic $\mathrm{HfO_2}$, where the off-grid MAE climbs to $91.56\text{ meV}$, whereas our basis holds it at $3.57\text{ meV}$. In AB-stacked bilayer graphene, whose low-lying conduction states extend into the vacuum, orthogonal mixing of a diffuse Gaussian component at fixed 13-orbital-per-atom dimension raises the minimum subspace projectability $p_{\min}$ from $0.394$ to $0.960$ and lowers the off-grid MAE to $5.18\text{ meV}$; transferring this carbon basis directly to 28-atom twisted bilayer graphene yields a band MAE of $4.23\text{ meV}$, as shown in Fig.~\ref{fig:bilayer_bands}(b).

\begin{table}[t]
\centering
\caption{Three-term off-grid band RMS error attribution (in meV) based on Eq.~\eqref{eq:three_term_eigen}.}
\label{tab:three_term_attribution}
\resizebox{\columnwidth}{!}{%
\begin{tabular}{llcccc}
\toprule
System & Source mesh & Basis & Interp. & Trunc. & Total \\
\midrule
$\beta\text{-}\mathrm{Ga_2O_3}$ & $3\times 3\times 3$ & $3.11$ & $2140.52$ & $<0.01$ & $2138.22$ \\
$\beta\text{-}\mathrm{Ga_2O_3}$ & $5\times 5\times 5$ & $3.06$ & $3.92$ & $0.01$ & $4.92$ \\
$\beta\text{-}\mathrm{Ga_2O_3}$ & $7\times 7\times 7$ & $3.10$ & $1.36$ & $0.10$ & $4.00$ \\
$\mathrm{GaAs}$ & $7\times 7\times 7$ & $10.00$ & $12.59$ & $<0.01$ & $15.72$ \\
$\mathrm{HfO_2}$ & $7\times 7\times 7$ & $1.65$ & $6.24$ & $0.05$ & $6.21$ \\
Monolayer $\mathrm{MoS_2}$ & $9\times 9\times 1$ & $2.31$ & $1.93$ & $<0.01$ & $3.27$ \\
$\mathrm{CsPbI_3}$ & $5\times 5\times 5$ & $1.32$ & $1.48$ & $<0.01$ & $1.57$ \\
\bottomrule
\end{tabular}}
\end{table}

Using the exact three-term decomposition in Eq.~\eqref{eq:three_term_eigen}, Fig.~\ref{fig:basis_attribution}(b) and Table~\ref{tab:three_term_attribution} cleanly separate the physical error contributions. The same-$\mathbf{k}$ finite-basis term $e_\star - E$ is independent of the source mesh density, directly establishing the intrinsic variational limit of the fixed AO space for the PAW operator. In contrast, the reciprocal-space interpolation term $e_d - e_\star$ decays rapidly upon mesh refinement---for example, dropping in $\beta\text{-}\mathrm{Ga_2O_3}$ from $2140.52\text{ meV}$ on a $3^3$ mesh to $3.92\text{ meV}$ ($5^3$) and $1.36\text{ meV}$ ($7^3$)---while real-space sparsity truncation remains below $0.1\text{ meV}$ across all cases.

Figures~\ref{fig:basis_attribution}(c) and~\ref{fig:bilayer_bands} highlight the advantage of our approach over finite-band projection tools (LOBSTER and PAOFLOW) and unfiltered direct operator reconstruction (HPRO) on identical VASP inputs. When conduction bands enter the target window, conventional finite-band projection is prone to strong level repulsion and band-gap collapse induced by low-projectability high-energy states; on perturbed Si and AB bilayer graphene, LOBSTER and PAOFLOW incur off-grid band errors of $24.6\text{--}135.6\text{ meV}$, whereas our direct PAW operator assembly with spectral-filtered Ritz regularization reduces the error to $6.2\text{--}7.2\text{ meV}$. Furthermore, the same-source comparison on 2H bilayer $\mathrm{MoS_2}$ in Fig.~\ref{fig:bilayer_bands}(d) reveals the variational basis bias of unfiltered direct projection: unfiltered HPRO exhibits an off-grid band RMS of $89.61\text{ meV}$ and a wavefunction residual of $1094\text{ meV}$. Applying our spectral filter directly to the HPRO operator lowers the band RMS to $3.54\text{ meV}$ and slashes the wavefunction residual ninefold to $121\text{ meV}$, while our full PAW spectral-filtered reconstruction achieves a more than 30-fold reduction in band RMS to $2.76\text{ meV}$ with a wavefunction residual of $221\text{ meV}$ (see Table~\ref{tab:supp_external} of the Supplemental Material).

\subsection{Magnetic, correlated, spin-orbit, and hybrid-functional systems}\label{sec:res_multifunctional}

\begin{figure}[t]
\centering
\includegraphics[width=\columnwidth]{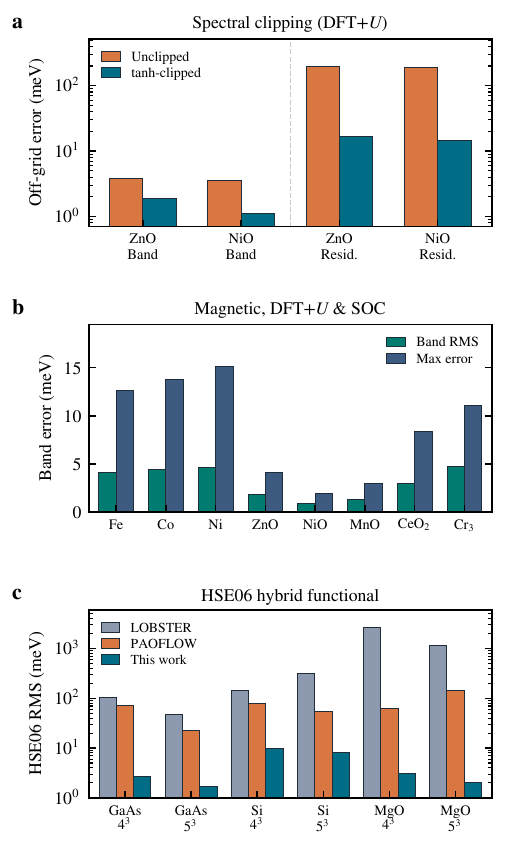}
\caption{(a) Effect of $\tanh$ spectral clipping [Eq.~\eqref{eq:tanh_clip}] on off-grid band errors and wavefunction residuals in DFT+$U$ $\mathrm{ZnO}$ and $\mathrm{NiO}$. (b) Off-grid band RMS and maximum errors across ferromagnetic metals, $3d/4f$ DFT+$U$ oxides, and noncollinear SOC $\mathrm{Cr_3}$. (c) HSE06 off-grid band RMS errors from LOBSTER, PAOFLOW, and our dual-mesh method.}
\label{fig:multifunctional}
\end{figure}

Figure~\ref{fig:multifunctional} summarizes our benchmarks on magnetic, strongly correlated, relativistic, and hybrid-functional VASP calculations. For DFT+$U$ oxides, Fig.~\ref{fig:multifunctional}(a) demonstrates the critical role of $\tanh$ high-energy spectral clipping [Eq.~\eqref{eq:tanh_clip}] in preserving wavefunction fidelity: without clipping, unresolved high-energy Ritz eigenvalues pushed upward by Hubbard $U$ amplify angular interpolation jitter, inflating wavefunction residuals in $\mathrm{ZnO}$ and $\mathrm{NiO}$ to $198.1$ and $190.6\text{ meV}$, respectively. Smooth spectral clipping halves the band RMS errors to $1.90$ and $1.11\text{ meV}$ while cutting the wavefunction residuals by more than a factor of 12 to $16.8$ and $14.7\text{ meV}$. Across $3d$ ferromagnetic metals ($\mathrm{Fe}$, $\mathrm{Co}$, $\mathrm{Ni}$), $3d$ DFT+$U$ oxides ($\mathrm{ZnO}$, antiferromagnetic $\mathrm{NiO}$, ferromagnetic $\mathrm{MnO}$), $4f$ $\mathrm{CeO_2}$, and noncollinear frustrated $\mathrm{Cr_3}$ with SOC [Fig.~\ref{fig:multifunctional}(b)], off-grid band RMS errors consistently remain within $0.96\text{--}4.79\text{ meV}$ (complete numerical results and wavefunction residuals are tabulated in Table~\ref{tab:supp_magnetic_dftu} of the Supplemental Material).

For VASP HSE06 screened hybrid-functional calculations on coarse $4^3$ and $5^3$ meshes, Fig.~\ref{fig:multifunctional}(c) shows that LOBSTER and PAOFLOW suffer severe off-grid band RMS errors ranging from $22.9$ to $2619.5\text{ meV}$, whereas single-mesh reconstruction either incurs larger errors ($5.4\text{--}49.7\text{ meV}$) or becomes rank-deficient in wide-gap $\mathrm{MgO}$ (Table~\ref{tab:supp_hse_comparison} of the Supplemental Material). Combining fine-grid semilocal anchoring [Eq.~\eqref{eq:hse_final_hr}] with window-first polar orthonormalization [Eq.~\eqref{eq:protected_polar}] supplies long-range semilocal hoppings from the fine PBE mesh while interpolating only the smooth nonlocal exchange correction on the coarse HSE06 mesh. This reduces off-grid band RMS errors across $\mathrm{GaAs}$, $\mathrm{Si}$, and wide-gap $\mathrm{MgO}$ to $1.71\text{--}8.03\text{ meV}$ on $5^3$ meshes and $2.74\text{--}9.97\text{ meV}$ even on coarse $4^3$ meshes---an improvement of one to three orders of magnitude.

\subsection{Graph neural-network learning and energy reference}\label{sec:res_transfer_mlh}

\begin{figure}[t]
\centering
\includegraphics[width=\columnwidth]{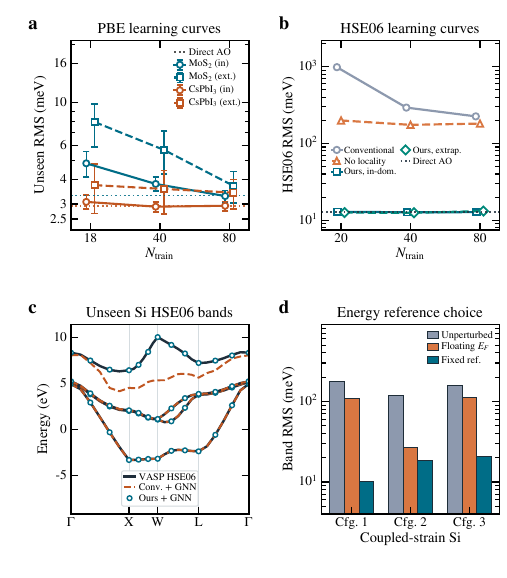}
\caption{(a) PBE GNN learning curves (95\% bootstrap intervals) for perturbed $\mathrm{MoS_2}$ and $\mathrm{CsPbI_3}$ relative to direct AO limits (dotted). (b) HSE06 $\mathrm{Si}$ learning curves comparing conventional, unregularized window-first, and locality-regularized labels. (c) Predicted versus VASP HSE06 band structure for an unseen extrapolated $\mathrm{Si}$ configuration ($N_{\mathrm{train}}=80$). (d) Coupled-strain $\mathrm{Si}$ band prediction errors under unperturbed, floating-$E_F$, and fixed electrostatic-potential references.}
\label{fig:transfer_gauge}
\end{figure}

To test whether the reconstructed sparse Hamiltonians $H_A(\mathbf{R})$ support smooth learning and stable extrapolation in realistic MLH pipelines, we constructed three 128-structure VASP datasets featuring six-degree-of-freedom lattice strains and internal atomic displacements: monolayer $\mathrm{MoS_2}$ (PBE), cubic $\mathrm{CsPbI_3}$ (PBE), and diamond $\mathrm{Si}$ (HSE06), totaling 384 independent self-consistent and off-grid reference calculations. Each dataset is partitioned into 80 training, 16 validation, and 24--32 held-out test structures. All models learn the Hamiltonian response $\Delta H_A(\mathbf{R})$ relative to the equilibrium structure; during inference on unseen configurations, only atomic coordinates are provided, and generalized eigenproblems are solved with purely geometric two-center overlaps $S_A(\mathbf{R})$ on 16--32 independent off-grid $\mathbf{k}$ points per test structure against raw VASP eigenvalues.

As shown in Fig.~\ref{fig:transfer_gauge}(a) (complete metrics in Table~\ref{tab:supp_mlh_full} of the Supplemental Material), equivariant HamGNN encoders paired with a response readout trained from scratch exhibit monotonic error reduction with training set size $N_{\mathrm{train}}$, converging directly to the underlying AO reconstruction limit. For $\mathrm{MoS_2}$, increasing $N_{\mathrm{train}}$ from 18 to 40 and 80 lowers the in-domain off-grid band RMS from $4.86$ to $3.80$ and $3.29\text{ meV}$---matching the $3.31\text{ meV}$ direct reconstruction limit---while outer-shell amplitude extrapolation errors drop from $7.95$ to $3.72\text{ meV}$. For $\mathrm{CsPbI_3}$, in-domain and extrapolation RMS errors reach $2.93$ and $3.43\text{ meV}$, with full real-space matrix-element MAEs of only $0.12\text{--}0.35\text{ meV}$ across both systems.

On the more demanding 128-structure VASP HSE06 dataset, relevant to hybrid-functional Hamiltonian learning~\cite{tang2024deephhybrid}, Figs.~\ref{fig:transfer_gauge}(b) and~\ref{fig:transfer_gauge}(c) demonstrate why both window-first polar orthonormalization and reciprocal-space locality regularization (Sec.~\ref{sec:locality_hse}) are indispensable for downstream machine learning. Evaluated across 24 prospective held-out structures, window-first labels lacking locality regularization stall GNN prediction errors at $174.8\text{--}199.5\text{ meV}$. Conventional labels improve from $985.8$ to $225.6\text{ meV}$ as $N_{\mathrm{train}}$ grows, yet suffer severe band collapse near the $W$ point in 11 of 24 test structures due to complement intrusion---illustrated by the $4.7\text{ eV}$ spurious band drop along the brown dashed curve at $W$ in Fig.~\ref{fig:transfer_gauge}(c). By contrast, our locality-regularized hybrid-functional labels hold GNN prediction errors within $12.5\text{--}13.2\text{ meV}$ across $N_{\mathrm{train}}=20,40,80$, achieving $12.76\text{ meV}$ (in-domain) and $13.17\text{ meV}$ (extrapolation) at $N_{\mathrm{train}}=80$ against a $12.75\text{ meV}$ direct AO limit, with a matrix-element MAE of $0.16\text{ meV}$. All 24 test structures remain completely free of $W$-point complement intrusion, and the $+0.63\text{--}0.71\text{ eV}$ HSE06-over-PBE band-gap opening is reproduced with a sampled gap RMS of $15.6\text{ meV}$ (full model comparisons are listed in Table~\ref{tab:supp_mlh_full} of the Supplemental Material).

Finally, Fig.~\ref{fig:transfer_gauge}(d) underscores the physical importance of a consistent cross-configuration energy reference, closely related to overlap-matrix energy-zero alignment in universal Hamiltonian models~\cite{wang2024universal}. On coupled volume-strain and internal-displacement Si test structures, subtracting each configuration's own finite-mesh nominal Fermi level to learn $H_A(\mathbf{R})-E_F S_A(\mathbf{R})$ introduces discontinuous shifts under volume deformation that degrade band predictions to $27.1\text{--}111.3\text{ meV}$. Preserving the common fixed VASP PAW electrostatic-potential reference and learning $H_A(\mathbf{R})$ directly reduces prediction errors to $9.1\text{--}20.8\text{ meV}$ (Table~\ref{tab:supp_si_transfer} of the Supplemental Material), underpinning the smooth cross-configuration learnability observed across all 384 structures.

\section{Conclusion}\label{sec:conclusion}

We have established a unified theoretical framework for reconstructing sparse effective Hamiltonians $H_A(\mathbf{R})$ in a fixed numerical atomic-orbital representation directly from PW-PAW calculations, and validated it systematically on VASP benchmarks. By decoupling the PAW projection metric $S_P(\mathbf{k})$ from the purely geometric two-center overlap $S_A(\mathbf{R})$, the framework retains exact PAW inner-product completeness during Hamiltonian extraction while enabling unseen-structure inference using only atomic coordinates and fixed compactly supported radial orbitals. To suppress variational basis bias and high-energy complement aliasing, we combined \emph{direct PAW operator assembly, spectral-filtered Ritz projection, and $C^2$ $\tanh$ high-energy spectral clipping} for semilocal, collinear magnetic, and DFT+$U$ systems with \emph{window-first two-step polar orthonormalization, reciprocal-space Voronoi smooth complement completion, and fine-grid semilocal-anchored dual-mesh interpolation} for noncollinear spinor, SOC, and HSE06 hybrid-functional calculations. Across diverse material benchmarks, the reconstructed Hamiltonians generally achieve millielectronvolt-scale off-grid band accuracy while substantially reducing wavefunction action residuals---by over an order of magnitude in strongly correlated oxides. End-to-end equivariant graph neural network benchmarks across 384 PBE and HSE06 perturbed structures confirm that the resulting sparse Hamiltonians exhibit smooth cross-configuration learnability under a consistent fixed VASP electrostatic-potential reference, converging monotonically to the direct reconstruction limit and providing a theoretical and methodological foundation for constructing high-accuracy, cross-functional MLH electronic-structure datasets.

\section*{Data and code availability}

The data and code that support the findings of this work are openly available at \href{https://doi.org/10.5281/zenodo.24465787}{doi:10.5281/zenodo.24465787}.

\begin{acknowledgments}
This work was supported by the National Natural Science Foundation of China (Grants No.~12374069 and No.~12674093) and the Science and Technology Program of Guangzhou (Grant No.~2025A04J5242). We thank the Center of Campus Network \& Modern Educational Technology, Guangdong University of Technology, Guangdong, China, for providing computational resources and technical support for this work.
\end{acknowledgments}

\bibliography{references}

\clearpage
\onecolumngrid

\setcounter{equation}{0}
\setcounter{figure}{0}
\setcounter{table}{0}
\setcounter{section}{0}
\renewcommand{\theequation}{S\arabic{equation}}
\renewcommand{\theHequation}{S\arabic{equation}}
\renewcommand{\thefigure}{S\arabic{figure}}
\renewcommand{\theHfigure}{S\arabic{figure}}
\renewcommand{\thetable}{S\arabic{table}}
\renewcommand{\theHtable}{S\arabic{table}}
\renewcommand{\thesection}{S\arabic{section}}
\renewcommand{\theHsection}{S\arabic{section}}

\makeatletter
\let\addtocontents\orig@addtocontents
\makeatother

\begin{center}
{\large\bfseries Supplemental Material: High-accuracy reconstruction of machine-learning Hamiltonians in a fixed atomic-orbital representation from plane-wave projector augmented-wave calculations\par}
\vspace{0.9em}
{\normalsize Qinghong Wu,$^{1,2}$ Jun Kang,$^{3}$ Minru Wen,$^{1,*}$ Huafeng Dong,$^{1}$ Le Huang,$^{1,2,\dagger}$ and Zhongming Wei$^{4,\ddagger}$\par}
\vspace{0.45em}
{\small\itshape
$^{1}$School of Physics and Optoelectronic Engineering, Guangdong University of Technology, Guangzhou 510006, China\\
$^{2}$School of Materials and Energy, Guangdong University of Technology, Guangzhou 510006, China\\
$^{3}$Beijing Computational Science Research Center, Beijing 100193, China\\
$^{4}$State Key Laboratory of Semiconductor Physics and Chip Technologies,\\
Institute of Semiconductors, Chinese Academy of Sciences, Beijing 100083, China\par}
\vspace{0.35em}
{\footnotesize $^{*}$\href{mailto:mrwen@gdut.edu.cn}{mrwen@gdut.edu.cn}\quad $^{\dagger}$\href{mailto:huangle@gdut.edu.cn}{huangle@gdut.edu.cn}\quad $^{\ddagger}$\href{mailto:zmwei@semi.ac.cn}{zmwei@semi.ac.cn}\par}
\end{center}

\vspace{0.6em}
{\linespread{0.95}\selectfont\tableofcontents}
\newpage

{\linespread{1.08}\selectfont

\section{Plane-wave projection, two-center overlap, and dual-metric congruence}\label{sec:supp_projection}

\subsection{Radial spherical Bessel transform}
In Eq.~\eqref{eq:pw_column} of the main text, expanding trial atomic orbitals $\mu=(a,\zeta,l,m)$ onto the plane-wave basis $\{\mathbf{q}=\mathbf{k}_{\mathrm{cart}}+\mathbf{G}\}$ satisfying $\frac{\hbar^2|\mathbf{k}_{\mathrm{cart}}+\mathbf{G}|^2}{2m_e}\le E_{\mathrm{cut}}$ requires evaluating the radial spherical Bessel transform as a function of momentum transfer magnitude $|\mathbf{q}|$:
\begin{equation}
I_{\zeta l}(|\mathbf{q}|) = \int_0^{r_c} r^2 u_{\zeta l}(r)\,j_l(|\mathbf{q}|r)\,\dd r.
\label{eq:supp_bessel}
\end{equation}
In our numerical quadrature, $j_l(x)$ is evaluated via power-series expansion at small arguments to avoid catastrophic cancellation and via explicit trigonometric recurrence at general arguments. Cubic spline interpolation over unique $|\mathbf{q}|$ values agrees with direct pointwise quadrature to within a maximum matrix-element difference below $1.1\times 10^{-11}\text{ eV}$ for both $H_A(\mathbf{R})$ and $S_A(\mathbf{R})$.

\subsection{Two-center overlap quadrature}
The real-space two-center atomic-orbital overlap matrix $S_A(\mathbf{R})$ defined in Eq.~\eqref{eq:sa_def} of the main text is evaluated separately for on-site and off-site blocks. For on-site blocks ($\mathbf{R}=\mathbf{0}$ and $a=b$), angular orthonormality of real spherical harmonics reduces the integral to $S_{A,(a,\zeta,l,m),(a,\zeta',l',m')}(\mathbf{0}) = \delta_{ll'}\delta_{mm'}\int_0^{r_c} r^2 u_{\zeta l}(r)u_{\zeta' l}(r)\,\dd r$. To avoid quadrature inconsistencies between diagonal normalization and off-diagonal radial overlaps, on-site blocks sharing a common radial grid reuse the exact Simpson quadrature rule employed to normalize $u_{\zeta l}(r)$, ensuring strict positive definiteness of $S_A(\mathbf{k})$.

For off-site blocks separated by the bond vector $\mathbf{d}=\bm{\tau}_b+\mathbf{R}\mathbf{a}-\bm{\tau}_a$ with $d=|\mathbf{d}|>0$, aligning the unit vector $\hat{\mathbf{d}}$ along the local $z$ axis and transforming to two-center cylindrical coordinates, $\dd^3\mathbf{r}=(rs/d)\,\dd r\,\dd s\,\dd\varphi$, reduces the axial overlap block to a two-dimensional integral diagonal in $|m|$:
\begin{equation}
S^z_{lm,l'm'}(d) = \delta_{mm'}\pi(1+\delta_{m0})
\iint_{\mathcal{D}(d)} \frac{rs}{d}\,u_a(r)u_b(s)\,F_{l|m|}(z_a)F_{l'|m|}(z_b)\,\dd s\,\dd r,
\label{eq:supp_two_center_axial}
\end{equation}
where $z_a=(r^2+d^2-s^2)/(2rd)$, $z_b=(r^2-d^2-s^2)/(2sd)$, $F_{l|m|}(z)$ denote normalized associated Legendre polynomials, and the integration domain is bounded by the orbital cutoffs as $\mathcal{D}(d)=\{(r,s)\mid 0\le r\le r_{c,a},\ |r-d|\le s\le\min(r+d,r_{c,b})\}$. Rotating back from the local bond axis to the global Cartesian frame via real spherical-harmonic rotation matrices $R_l(\hat{\mathbf{d}})$ yields $S_{A,\mu\nu}(\mathbf{R}) = [R_l^T(\hat{\mathbf{d}}) S^z(d) R_{l'}(\hat{\mathbf{d}})]_{mm'}$. Doubling the piecewise Gauss--Legendre order from $96$ to $192$ on AB-stacked bilayer graphene alters off-site $S_A(\mathbf{R})$ elements by at most $1.33\times 10^{-4}$ and shifts four-band eigenvalues across 96 off-grid $\mathbf{k}$ points by only $0.024\text{ meV}$ RMS.

\subsection{PAW eigenproblem and dual-metric congruence}\label{sec:supp_dual_metric}

In Bl\"ochl's PAW formalism~\cite{blochl1994}, the all-electron Kohn--Sham state $|\Psi_{n\mathbf{k}}\rangle$ is related to the smooth pseudo-wavefunction $|\widetilde{\Psi}_{n\mathbf{k}}\rangle$ by the invertible linear transformation
\begin{equation}
\mathcal{T} = 1 + \sum_{a,i}\bigl(|\phi_{ai}^{\mathrm{AE}}\rangle - |\tilde{\phi}_{ai}^{\mathrm{PS}}\rangle\bigr)\langle \tilde{p}_{ai}|,
\label{eq:supp_paw_T}
\end{equation}
where $|\phi_{ai}^{\mathrm{AE}}\rangle$, $|\tilde{\phi}_{ai}^{\mathrm{PS}}\rangle$, and $|\tilde{p}_{ai}\rangle$ denote the all-electron partial waves, pseudo partial waves, and dual projector functions localized within the augmentation sphere of atom $a$. Substituting $|\Psi_{n\mathbf{k}}\rangle = \mathcal{T}|\widetilde{\Psi}_{n\mathbf{k}}\rangle$ into the all-electron Kohn--Sham equation $H^{\mathrm{AE}}|\Psi_{n\mathbf{k}}\rangle = E_{n\mathbf{k}}|\Psi_{n\mathbf{k}}\rangle$ and left-multiplying by $\mathcal{T}\tr$ yields the PAW generalized eigenproblem in the plane-wave representation of the smooth pseudo-wavefunctions:
\begin{equation}
\widetilde{H}(\mathbf{k})\,C(\mathbf{k}) = O(\mathbf{k})\,C(\mathbf{k})\,E(\mathbf{k}),\qquad C\tr(\mathbf{k})\,O(\mathbf{k})\,C(\mathbf{k}) = I_{N_b},
\label{eq:supp_paw_ks}
\end{equation}
where $\widetilde{H}(\mathbf{k}) = \mathcal{T}\tr H^{\mathrm{AE}}\mathcal{T}$ is the pseudo-Hamiltonian and $O(\mathbf{k}) = \mathcal{T}\tr\mathcal{T} = I + P(\mathbf{k})\Delta P\tr(\mathbf{k})$ is the PAW overlap operator with one-center augmentation overlap tensor $\Delta_{ai,aj} = \langle\phi_{ai}^{\mathrm{AE}}|\phi_{aj}^{\mathrm{AE}}\rangle - \langle\tilde{\phi}_{ai}^{\mathrm{PS}}|\tilde{\phi}_{aj}^{\mathrm{PS}}\rangle$. Because $\mathcal{T}$ is an invertible map from smooth pseudo-wavefunctions to all-electron wavefunctions under local partial-wave completeness, $O(\mathbf{k}) = \mathcal{T}\tr\mathcal{T}\succ 0$ is strictly Hermitian positive definite (HPD).

Projecting onto the fixed $N_{\mathrm{AO}}$-dimensional trial atomic-orbital basis $\{\phi_{\mu\mathbf{R}}\}_{\mu=1}^{N_{\mathrm{AO}}}$ introduces two distinct $N_{\mathrm{AO}}\times N_{\mathrm{AO}}$ Hermitian metrics on the identical orbital index set. The first is the PAW projection metric $S_P(\mathbf{k}) = \Phi\tr(\mathbf{k}) O(\mathbf{k})\Phi(\mathbf{k}) \in \mathbb{C}^{N_{\mathrm{AO}}\times N_{\mathrm{AO}}}$; because $O(\mathbf{k})\succ 0$ and the plane-wave expansion matrix $\Phi(\mathbf{k})\in\mathbb{C}^{N_{\mathrm{PW}}(\mathbf{k})\times N_{\mathrm{AO}}}$ has full column rank $N_{\mathrm{AO}} \ll N_{\mathrm{PW}}(\mathbf{k})$, every nonzero vector $v\in\mathbb{C}^{N_{\mathrm{AO}}}$ satisfies $v\tr S_P(\mathbf{k}) v = (\mathcal{T}\Phi v)\tr(\mathcal{T}\Phi v) > 0$, rendering $S_P(\mathbf{k})\succ 0$ strictly HPD. The second is the continuum geometric AO metric $S_A(\mathbf{k}) = \sum_{\mathbf{R}} e^{2\pi i \mathbf{k}\cdot\mathbf{R}} S_A(\mathbf{R}) \in \mathbb{C}^{N_{\mathrm{AO}}\times N_{\mathrm{AO}}}$, which represents the exact continuum Gram matrix $\langle\phi_{\mu\mathbf{k}}|\phi_{\nu\mathbf{k}}\rangle_{L^2(\mathbb{R}^3)}$ of the Bloch-summed compact atomic orbitals without plane-wave cutoff truncation or PAW augmentation. Because the radial basis is Gram--Schmidt orthonormalized within each $(Z,l)$ channel with linear-independence threshold $>10^{-8}$ (Sec.~\ref{sec:supp_basis}) and the compact cutoff $r_c$ maintains a bounded condition number $\kappa_{\max}(S_A)$ throughout the Brillouin zone, $S_A(\mathbf{k})\succ 0$ is likewise strictly HPD for all $\mathbf{k}$. In general $S_P(\mathbf{k}) \neq S_A(\mathbf{k})$, since $S_P(\mathbf{k}) = \Phi\tr\Phi + \Phi\tr P\Delta P\tr\Phi$ incorporates both finite-$E_{\mathrm{cut}}$ plane-wave truncation ($\Phi\tr\Phi\neq S_A(\mathbf{k})$) and the one-center PAW augmentation overlap $\Phi\tr P\Delta P\tr\Phi$.

To convert rigorously from the PAW metric $S_P(\mathbf{k})$ to the purely geometric AO metric $S_A(\mathbf{k})$ without spectral distortion, we exploit the unique HPD principal square roots $S_P(\mathbf{k})^{1/2}\succ 0$ and $S_A(\mathbf{k})^{1/2}\succ 0$ to define the invertible L\"owdin metric-transfer operator
\begin{equation}
M(\mathbf{k}) \equiv S_A(\mathbf{k})^{1/2}\,S_P(\mathbf{k})^{-1/2} \in \mathbb{C}^{N_{\mathrm{AO}}\times N_{\mathrm{AO}}}, \qquad M(\mathbf{k})\,S_P(\mathbf{k})\,M\tr(\mathbf{k}) = S_A(\mathbf{k}).
\label{eq:supp_metric_transfer}
\end{equation}
Denoting by $h(\mathbf{k}) = h\tr(\mathbf{k}) \in \mathbb{C}^{N_{\mathrm{AO}}\times N_{\mathrm{AO}}}$ the effective Hamiltonian constructed in the common L\"owdin-orthonormalized frame [via either Eqs.~\eqref{eq:filtered_h}--\eqref{eq:tanh_clip} or Eq.~\eqref{eq:voronoi_objective} of the main text], the corresponding PAW-metric and AO-metric Hamiltonians are related by the congruence transformation
\begin{equation}
H_P^{\mathrm{eff}}(\mathbf{k}) = S_P(\mathbf{k})^{1/2}\,h(\mathbf{k})\,S_P(\mathbf{k})^{1/2}, \qquad
H_A(\mathbf{k}) = S_A(\mathbf{k})^{1/2}\,h(\mathbf{k})\,S_A(\mathbf{k})^{1/2} = M(\mathbf{k})\,H_P^{\mathrm{eff}}(\mathbf{k})\,M\tr(\mathbf{k}).
\label{eq:supp_congruence}
\end{equation}
Consequently, the generalized eigenproblems $H_P^{\mathrm{eff}}(\mathbf{k}) c_P = S_P(\mathbf{k}) c_P \Lambda$ and $H_A(\mathbf{k}) c_A = S_A(\mathbf{k}) c_A \Lambda$ share the exact dimension $N_{\mathrm{AO}}\times N_{\mathrm{AO}}$ and an identical eigenvalue spectrum $\Lambda(\mathbf{k})$, while their generalized eigenvectors are linked bijectively through $c_A(\mathbf{k}) = M^{-\dagger}(\mathbf{k})\,c_P(\mathbf{k}) =\allowbreak S_A(\mathbf{k})^{-1/2}S_P(\mathbf{k})^{1/2}c_P(\mathbf{k})$, preserving generalized orthonormality:
\begin{equation}
c_A\tr(\mathbf{k}) S_A(\mathbf{k}) c_A(\mathbf{k}) = c_P\tr(\mathbf{k}) S_P(\mathbf{k}) c_P(\mathbf{k}) = X\tr(\mathbf{k}) X(\mathbf{k}) = I_{N_{\mathrm{AO}}},
\label{eq:supp_iso_proof}
\end{equation}
where $X(\mathbf{k}) = S_P(\mathbf{k})^{1/2} c_P(\mathbf{k}) = S_A(\mathbf{k})^{1/2} c_A(\mathbf{k})$ is the common orthonormal eigenbasis of $h(\mathbf{k})$. Moreover, unlike asymmetric Cholesky factorizations that depend on orbital ordering and break rotational equivariance, the HPD principal square roots $S_P(\mathbf{k})^{1/2}$ and $S_A(\mathbf{k})^{1/2}$ commute with every orthogonal real-spherical-harmonic representation $D(\mathcal{R})$ ($\mathcal{R}\in O(3)$), so $D(\mathcal{R})S_A(\mathbf{k})^{1/2}D\tr(\mathcal{R}) =\allowbreak S_A(\mathcal{R}\mathbf{k})^{1/2}$ and identically for $S_P(\mathbf{k})^{1/2}$. The congruence map~\eqref{eq:supp_congruence} therefore strictly preserves $E(3)$ equivariance of $H_A(\mathbf{R})$ for downstream equivariant neural networks while completely eliminating the first-order metric-mismatch error $\delta\Lambda \sim X\tr(S_A^{-1/2}S_P S_A^{-1/2} - I)X\Lambda$ that would arise if $H_P^{\mathrm{eff}}(\mathbf{k})$ were paired directly with $S_A(\mathbf{k})$.

\section{Compact numerical atomic-orbital construction}\label{sec:supp_basis}

For species $Z$ with cutoff radius $r_c$ and angular-momentum multiplicities $n_{Zl}$, the single-atom spatial orbital dimension is $d_Z=\sum_l (2l+1)n_{Zl}$. On a 513-point uniform radial grid over $[0,r_c]$, we define the $C^2$-continuous quintic smooth taper
\begin{equation}
f(r) = 1 - 10t^3 + 15t^4 - 6t^5,\qquad t = \mathrm{clip}\!\left(\frac{r/r_c - 0.7}{0.3},\,0,\,1\right).
\label{eq:supp_taper_func}
\end{equation}
Candidate radial functions $f(r)u_{il}^{\mathrm{PAW}}(r)/r$ are first extracted from the native VASP PAW atomic reference datasets. Whenever $n_{Zl}$ exceeds the number of bound partial waves or additional polarization shells are required, compactly supported Slater-type radial functions with geometrically decaying exponents are appended sequentially:
\begin{equation}
b_{lj}(r) = f(r)\left(\frac{r}{r_c}\right)^l \exp\!\left[-(l+2)\,1.6^j\,\frac{r}{r_c}\right],\qquad j=0,1,\ldots
\label{eq:supp_slater_complement}
\end{equation}
Within each $l$ channel, two-pass modified Gram--Schmidt orthonormalization is performed using the Simpson inner product $\langle u_i|u_j\rangle = \int_0^{r_c} r^2 u_i(r)u_j(r)\,\dd r$, discarding nearly linearly dependent directions whose squared residual falls below $10^{-8}$ of its initial norm and fixing the sign of each accepted radial orbital to be positive at the peak of $|r u(r)|$. For two-dimensional slabs with vacuum-extended conduction states, outer shells are further mixed orthogonally with a diffuse Gaussian trial function $g_l(r;\alpha_g)\propto r^l e^{-r^2/(2\alpha_g^2)}f(r)$ chosen to maximize the minimum subspace projectability $p_{\min}=\min_{\mathbf{k}}\lambda_{\min}[V_e\tr(\mathbf{k})V_e(\mathbf{k})]$ of the orthonormalized target-window projection $V_e(\mathbf{k})=S_P(\mathbf{k})^{-1/2}A_e(\mathbf{k})$.

\section{Spectral filtering, complement completion, and operator diagnostics}\label{sec:supp_derivations}

\subsection{Spectral-filtered Ritz projection and high-energy clipping}\label{sec:supp_filter_proof}

Let $\hat{h}_\sigma(\mathbf{k}) = O(\mathbf{k})^{-1/2}\widetilde{H}_\sigma(\mathbf{k})O(\mathbf{k})^{-1/2}$ be the self-adjoint PAW Hamiltonian on the full plane-wave space $\mathcal{H}_{\mathrm{PW}}(\mathbf{k})$ of dimension $N_{\mathrm{PW}}(\mathbf{k})$, with complete orthonormal eigenbasis $\{|\tilde{u}_n\rangle = O^{1/2}C_{n\sigma}\}_{n=1}^{N_{\mathrm{PW}}}$ and ordered eigenvalues $E_{1\mathbf{k}\sigma}\le \cdots \le E_{N_{\mathrm{PW}}\mathbf{k}\sigma}$, of which only the lowest $N_b \ll N_{\mathrm{PW}}$ bands are explicitly computed. Define the PAW-orthonormalized trial AO basis $\Xi(\mathbf{k}) = O(\mathbf{k})^{1/2}\Phi(\mathbf{k})S_P(\mathbf{k})^{-1/2} \in \mathbb{C}^{N_{\mathrm{PW}}\times N_{\mathrm{AO}}}$, which satisfies $\Xi\tr\Xi = I_{N_{\mathrm{AO}}}$ and $\Xi\tr\hat{h}_\sigma\Xi = S_P^{-1/2}H_{P,\sigma}S_P^{-1/2} \equiv K_{0,\sigma}(\mathbf{k})$. The orthonormalized band-projection matrix is $V_\sigma(\mathbf{k}) = S_P(\mathbf{k})^{-1/2}A_\sigma(\mathbf{k}) = \Xi\tr U_{1:N_b}$, where $U_{1:N_b} = O^{1/2}C_\sigma$ satisfies $U_{1:N_b}\tr U_{1:N_b} = I_{N_b}$.

The smooth filter weights $F(E)^2 = \beta + w(E)$ in Eq.~\eqref{eq:filter_weight} of the main text are governed by physical energy scales referenced to the Fermi level $E_F$ and the target-window upper bound $E_{\mathrm{win}}^{\max}$ (typically $E_F+2$ to $E_F+4\text{ eV}$). Specifically, we set the filter center to $E_c = \max(E_F + 6\text{ eV},\, E_{\mathrm{win}}^{\max} + 2\text{ eV})$ with Fermi--Dirac transition width $\Delta_b = 1.0\text{ eV}$, ensuring $\{1+\exp[(E-E_c)/\Delta_b]\}^{-1} > 0.88$ throughout the target window $E\le E_{\mathrm{win}}^{\max}$ and $<0.05$ for $E > E_c + 3\Delta_b$. Above the target window, the upper band-cutoff floor $E_t = \min_{\mathbf{k}} E_{N_b\mathbf{k}\sigma}$ is paired with a $C^1$ smoothstep taper width $\Delta_d = 2.0\text{ eV}$, enforcing $w(E)=0$ and $w'(E)=0$ identically for all $E\ge E_t$ so that band crossings across the highest computed band $N_b$ introduce no reciprocal-space discontinuities. Setting the regularization floor to $\beta = 0.005$ balances a $200$-fold suppression of high-energy variational admixture against a globally bounded overlap condition number $\kappa_2(G_\sigma)\le \beta^{-1} = 200$, while the subsequent spectral clipping map uses onset $E_{\mathrm{clip}} = E_{\mathrm{win}}^{\max} + 2\text{ eV}$ and width $\Delta_{\mathrm{clip}} = 2.0\text{ eV}$.

Because $w(E_n) = 0$ identically for every uncomputed plane-wave eigenstate $n > N_b$ ($E_{n\mathbf{k}\sigma}\ge E_{N_b\mathbf{k}\sigma}\ge E_t$), inserting the complete plane-wave resolution of the identity $\sum_{n=1}^{N_{\mathrm{PW}}} |\tilde{u}_n\rangle\langle\tilde{u}_n| = I_{\mathcal{H}_{\mathrm{PW}}}$ folds the entire infinite orthogonal complement into the $\beta$ floor without truncation:
\begin{align}
G_\sigma(\mathbf{k}) &\equiv \Xi\tr F(\hat{h}_\sigma)^2 \Xi = \beta \Xi\tr\Xi + \sum_{n=1}^{N_b} w_{n\sigma} (\Xi\tr|\tilde{u}_n\rangle)(\langle\tilde{u}_n|\Xi) \nonumber\\
&= \beta I_{N_{\mathrm{AO}}} + V_\sigma \mathrm{diag}(w_{n\sigma}) V_\sigma\tr, \label{eq:supp_G_closure}\\
K_\sigma(\mathbf{k}) &\equiv \Xi\tr F(\hat{h}_\sigma)\hat{h}_\sigma F(\hat{h}_\sigma)\Xi = \beta \Xi\tr\hat{h}_\sigma\Xi + \sum_{n=1}^{N_b} w_{n\sigma}E_{n\mathbf{k}\sigma} (\Xi\tr|\tilde{u}_n\rangle)(\langle\tilde{u}_n|\Xi) \nonumber\\
&= \beta K_{0,\sigma} + V_\sigma \mathrm{diag}(w_{n\sigma}E_{n\mathbf{k}\sigma}) V_\sigma\tr. \label{eq:supp_K_closure}
\end{align}
Since $0 \le w_{n\sigma} \le 1-\beta$ and $0\preceq V_\sigma V_\sigma\tr = \Xi\tr U_{1:N_b}U_{1:N_b}\tr\Xi \preceq \Xi\tr\Xi = I_{N_{\mathrm{AO}}}$, the filtered overlap moment obeys the uniform spectral bounds
\begin{equation}
\beta I_{N_{\mathrm{AO}}} \preceq G_\sigma(\mathbf{k}) \preceq I_{N_{\mathrm{AO}}}, \qquad \kappa_2\bigl(G_\sigma(\mathbf{k})\bigr) \le \beta^{-1} = 200.
\label{eq:supp_G_bounds}
\end{equation}
Consequently, the filtered trial subspace $\widetilde{\Xi} = F(\hat{h}_\sigma)\Xi$ retains full column rank $N_{\mathrm{AO}}$ with no discarded orbital dimensions, and symmetric L\"owdin orthonormalization $\widehat{\Xi} = \widetilde{\Xi}\,G_\sigma(\mathbf{k})^{-1/2}$ remains globally well-conditioned, satisfies $\widehat{\Xi}\tr\widehat{\Xi}=I_{N_{\mathrm{AO}}}$, and defines the Rayleigh--Ritz operator $h_\sigma(\mathbf{k}) = \widehat{\Xi}\tr\hat{h}_\sigma\widehat{\Xi} = G_\sigma^{-1/2}K_\sigma G_\sigma^{-1/2}$.

This construction strictly preserves target-space spanning while suppressing variational bias. For any target eigenstate $|\tilde{u}_m\rangle$ contained in the trial span ($|\tilde{u}_m\rangle = \Xi v_m$ with $\|v_m\|_2=1$), $F(\hat{h}_\sigma)|\tilde{u}_m\rangle = F(E_m)|\tilde{u}_m\rangle \in \mathrm{span}(\widetilde{\Xi})$, so $|\tilde{u}_m\rangle$ remains an exact eigenstate of $h_\sigma(\mathbf{k})$ with eigenvalue $E_m$ for every $\beta>0$; when $|\tilde{u}_m\rangle$ carries a small high-energy orthogonal leakage component above $E_c$, $F(\hat{h}_\sigma)$ attenuates that component in $\Xi$ by $\sqrt{\beta}$ relative to $F(E_m)\approx 1$, reducing the squared high-energy variational admixture by a factor of $\beta = 0.005$. Likewise, the subsequent $C^2$ clipping map $f_{\mathrm{clip}}(h_\sigma) = X_\sigma f_{\mathrm{clip}}(\Lambda_\sigma) X_\sigma\tr$ [Eq.~\eqref{eq:tanh_clip} of the main text] acts purely spectrally on the unchanged complete orthonormal eigenbasis $X_\sigma(\mathbf{k})$ ($X_\sigma\tr X_\sigma = I_{N_{\mathrm{AO}}}$). Because $f_{\mathrm{clip}}(x) = x$ for all $x\le E_{\mathrm{clip}}$ with $f'_{\mathrm{clip}}(E_{\mathrm{clip}}^+)=1$, $f''_{\mathrm{clip}}(E_{\mathrm{clip}}^+)=0$, and $f'_{\mathrm{clip}}(x)=\mathrm{sech}^2[(x-E_{\mathrm{clip}})/\Delta_{\mathrm{clip}}]>0$ for $x>E_{\mathrm{clip}}$, every eigenvector, every eigenvalue below $E_{\mathrm{clip}}$, and the global eigenvalue ordering are strictly preserved, whereas unresolved high-energy modes during Fourier interpolation are bounded above by $E_{\mathrm{clip}}+\Delta_{\mathrm{clip}}$.

In numerical implementation for semilocal, collinear magnetic, and DFT+$U$ calculations, each spin channel $\sigma$ and source wave vector $\mathbf{k}\in K_{\mathrm{src}}$ is processed sequentially by evaluating $\Phi(\mathbf{k})$ via Eq.~\eqref{eq:pw_column} of the main text, assembling $S_P(\mathbf{k}) = \Phi\tr O\Phi$, $A_\sigma(\mathbf{k}) = \Phi\tr O C_\sigma$, and $H_{P,\sigma}(\mathbf{k})$ via Eqs.~\eqref{eq:direct_paw_action}--\eqref{eq:hp_def}, and transforming to the PAW-orthonormalized representations $V_\sigma(\mathbf{k}) = S_P^{-1/2}A_\sigma$ and $K_{0,\sigma}(\mathbf{k}) = S_P^{-1/2}H_{P,\sigma}S_P^{-1/2}$. After evaluating the smooth energy weights $w_{n\sigma}$ from Eq.~\eqref{eq:filter_weight}, the moments $G_\sigma(\mathbf{k})$ and $K_\sigma(\mathbf{k})$ are formed via Eqs.~\eqref{eq:supp_G_closure}--\eqref{eq:supp_K_closure} to obtain $h_\sigma(\mathbf{k}) = G_\sigma^{-1/2}K_\sigma G_\sigma^{-1/2}$ and its eigendecomposition $h_\sigma(\mathbf{k}) = X_\sigma\Lambda_\sigma X_\sigma\tr$. Applying $f_{\mathrm{clip}}(\Lambda_\sigma)$ and pulling back through $S_A(\mathbf{k})^{1/2}$ via Eq.~\eqref{eq:ha_pullback} yields $H_{A,\sigma}(\mathbf{k}) = S_A(\mathbf{k})^{1/2} X_\sigma f_{\mathrm{clip}}(\Lambda_\sigma) X_\sigma\tr S_A(\mathbf{k})^{1/2}$, from which the sparse real-space Hamiltonian $H_{A,\sigma}(\mathbf{R})$ is obtained through the energy-origin-covariant Fourier transform of Eqs.~\eqref{eq:offset_dft}--\eqref{eq:gauge_covariant_hr} with $\eta=E_F$.

\subsection{Window-protected polar orthonormalization and complement completion}\label{sec:supp_voronoi_proof}

For noncollinear spinor, SOC, and HSE06 hybrid-functional calculations, where an explicit local-potential operator action $\widetilde{H}\Phi$ is unavailable or involves nonlocal exact exchange, the orthogonal complement is reconstructed by solving a strictly convex constrained optimization problem. At each $\mathbf{k}\in K_{\mathrm{src}}$, let $V_T\in\mathbb{C}^{N_{\mathrm{AO}}\times t}$ denote the PAW-orthonormalized target-window projection columns ($\sigma_{\min}(V_T)=\sqrt{p_{\min}}>0$) and $V_U\in\mathbb{C}^{N_{\mathrm{AO}}\times (m_{\mathbf{k}}-t)}$ the high-energy buffer columns ($m_{\mathbf{k}}=t+2$, extended locally to complete any degenerate multiplet within $10^{-4}\text{ eV}$). Forming $Q_T = \mathrm{polar}(V_T) = V_T(V_T\tr V_T)^{-1/2}$ yields the unique isometric minimizer of $\min_{Q\tr Q=I_t}\norm{Q-V_T}_F$ with $\mathrm{col}(Q_T) = \mathrm{col}(V_T)$ identically; deflating the buffer columns onto $\mathrm{col}(Q_T)^\perp$ via $\widetilde{V}_U = (I - Q_T Q_T\tr)V_U$ ($\sigma_{\min}(\widetilde{V}_U) > 10^{-6}$) and setting $Q_U = \mathrm{polar}(\widetilde{V}_U)$ then ensures $Q_T\tr Q_U = 0$ and $Q_U\tr Q_U = I_{m_{\mathbf{k}}-t}$, so the band-ordered frame $Q(\mathbf{k}) = [Q_T, Q_U]$ satisfies $Q\tr(\mathbf{k})Q(\mathbf{k}) = I_{m_{\mathbf{k}}}$ while strictly protecting $\mathrm{col}(Q_T)$ from buffer-induced rotations. Introducing any orthonormal basis $Z(\mathbf{k})\in\mathbb{C}^{N_{\mathrm{AO}}\times (N_{\mathrm{AO}}-m_{\mathbf{k}})}$ of $\mathrm{col}(Q(\mathbf{k}))^\perp$ ($Z\tr Z=I$, $Q\tr Z=0$, $QQ\tr + ZZ\tr = I_{N_{\mathrm{AO}}}$), every Hermitian matrix $h(\mathbf{k})$ satisfying Eq.~\eqref{eq:feasible_set} of the main text admits the unique block-diagonal representation
\begin{equation}
h(\mathbf{k}) = Q(\mathbf{k})\,E_{1:m_{\mathbf{k}}}(\mathbf{k})\,Q\tr(\mathbf{k}) + Z(\mathbf{k})\,h_\perp(\mathbf{k})\,Z\tr(\mathbf{k}), \qquad h_\perp(\mathbf{k}) = h_\perp\tr(\mathbf{k}) \succeq \bigl(E_{m_{\mathbf{k}}\mathbf{k}} + \Delta_c\bigr)I.
\label{eq:supp_block_h}
\end{equation}
Every target-window state in $Q_T(\mathbf{k})$ is therefore an exact eigenvector of $h(\mathbf{k})$ with exact plane-wave eigenvalue $E_{1:t}(\mathbf{k})$, completely decoupled from the complement block $h_\perp(\mathbf{k})$, whose spectrum is bounded below by $E_{m_{\mathbf{k}}\mathbf{k}} + \Delta_c > E_{t\mathbf{k}}$ ($\Delta_c = 0.1\text{ eV}$).

The product feasible set $\mathcal{C} = \prod_{\mathbf{k}\in K_{\mathrm{src}}}\mathcal{C}_{\mathbf{k}}$ defined by Eq.~\eqref{eq:supp_block_h} forms a closed convex subset of Hermitian matrix fields isomorphic to a Cartesian product of shifted positive-semidefinite cones. For any regularization weight $\gamma > 0$, the Hessian of the Voronoi smoothness functional $\mathcal{J}_{\mathrm{loc}}$ [Eq.~\eqref{eq:voronoi_objective} of the main text] satisfies $\nabla^2\mathcal{J}_{\mathrm{loc}} = (1-\gamma)L_{\mathrm{Vor}} + \gamma I \succeq \gamma I \succ 0$, where $L_{\mathrm{Vor}}\succeq 0$ denotes the gauge-covariant reciprocal-space graph Laplacian. Thus $\mathcal{J}_{\mathrm{loc}}$ is $\gamma$-strongly convex with Lipschitz gradient constant $L_{\mathrm{grad}} = \gamma + (1-\gamma)\sum_{\mathbf{b}} w_{\mathbf{b}} \le 2-\gamma < 2$, guaranteeing a unique global minimizer $\{h^\star(\mathbf{k})\}\in\mathcal{C}$ independent of initialization. Moreover, for any Hermitian iterate $\tilde{h}(\mathbf{k})$, its unique Frobenius-norm Euclidean projection onto $\mathcal{C}_{\mathbf{k}}$ is given in closed form by eigendecomposing $W = Z\tr\tilde{h}Z = U_W\mathrm{diag}(\omega_i)U_W\tr$, flooring $\tilde{\omega}_i = \max(\omega_i, E_{m_{\mathbf{k}}\mathbf{k}} + \Delta_c)$, and assembling
\begin{equation}
\Pi_{\mathcal{C}_{\mathbf{k}}}(\tilde{h}) = Q\,E_{1:m_{\mathbf{k}}}\,Q\tr + Z\,U_W\,\mathrm{diag}(\tilde{\omega}_i)\,U_W\tr\,Z\tr.
\label{eq:supp_proj_step}
\end{equation}
Because Eq.~\eqref{eq:supp_proj_step} depends solely on the orthogonal complement projector $P_\perp(\mathbf{k}) = ZZ\tr = I - QQ\tr$ and is invariant under arbitrary unitary rotations $Z\mapsto Z U_Z$, both the optimization trajectory and the unique global minimizer $h^\star(\mathbf{k})$ are strictly independent of the choice of complement basis $Z(\mathbf{k})$ and converge linearly at rate $O\bigl((1-\gamma/L_{\mathrm{grad}})^r\bigr)$ for step size $\alpha \le 1/L_{\mathrm{grad}}$.

In numerical execution, after forming $V(\mathbf{k}) = S_P(\mathbf{k})^{-1/2}A(\mathbf{k})$ at each $\mathbf{k}\in K_{\mathrm{src}}$ [preceded for spinor/SOC systems by the $SU(2)$ spin-axis alignment of Eq.~\eqref{eq:supp_su2_saxis} in $2N_{\mathrm{AO}}$ space] and assembling $Q(\mathbf{k})$ via Eq.~\eqref{eq:protected_polar} of the main text, we specify the reference baseline $h_{\mathrm{ref}}(\mathbf{k}) = \eta_0 I$ ($\eta_0 = E_F + 20\text{ eV}$) for single-mesh noncollinear/SOC reconstructions, or $h_{\mathrm{ref}}(\mathbf{k}) = h^{\mathrm{anc}}(\mathbf{k}) =\allowbreak S_A(\mathbf{k})^{-1/2}H_A^{\mathrm{anc}}(\mathbf{k})S_A(\mathbf{k})^{-1/2}$ obtained from a fine-mesh PBE spectral-filtered reconstruction (Sec.~\ref{sec:supp_filter_proof}) for dual-mesh HSE06 calculations. Starting from $h^{(0)}(\mathbf{k}) = \Pi_{\mathcal{C}_{\mathbf{k}}}\bigl(h_{\mathrm{ref}}(\mathbf{k})\bigr)$, projected gradient descent is iterated on the difference field $\Delta h^{(r)}(\mathbf{k}) = h^{(r)}(\mathbf{k}) - h_{\mathrm{ref}}(\mathbf{k})$ according to
\begin{align}
g^{(r)}(\mathbf{k}) &= \frac{1-\gamma}{2}\sum_{\mathbf{b}} w_{\mathbf{b}}\bigl[\Delta h^{(r)}(\mathbf{k}) - \Theta_{\mathbf{b}}\tr \Delta h^{(r)}(\mathbf{k}+\mathbf{b})\Theta_{\mathbf{b}}\bigr] + \gamma\,\Delta h^{(r)}(\mathbf{k}), \label{eq:supp_pgd_grad}\\
h^{(r+1)}(\mathbf{k}) &= \Pi_{\mathcal{C}_{\mathbf{k}}}\!\bigl(h^{(r)}(\mathbf{k}) - \alpha\, g^{(r)}(\mathbf{k})\bigr), \label{eq:supp_pgd}
\end{align}
with $\gamma=0.01$ and $\alpha = 1/L_{\mathrm{grad}}$ until $\max_{\mathbf{k}}\norm{h^{(r+1)}(\mathbf{k}) - h^{(r)}(\mathbf{k})}_F < 10^{-8}\text{ eV}$. At convergence, single-mesh calculations pull back $H_A(\mathbf{k}) = S_A(\mathbf{k})^{1/2} h^\star(\mathbf{k}) S_A(\mathbf{k})^{1/2}$ and apply Eqs.~\eqref{eq:offset_dft}--\eqref{eq:gauge_covariant_hr} of the main text, whereas dual-mesh HSE06 calculations transform only the smooth coarse-grid correction $S_A(\mathbf{k})^{1/2}\Delta h^\star(\mathbf{k})S_A(\mathbf{k})^{1/2}$ on $K_{\mathrm{HSE}}$ and superpose it onto $H_A^{\mathrm{anc}}(\mathbf{R})$ via Eq.~\eqref{eq:hse_final_hr} of the main text.

\subsection{Operator consistency and split-mesh reconstruction}
Prior to projecting the direct PAW operator [Eqs.~\eqref{eq:direct_paw_action}--\eqref{eq:hp_def} of the main text], operator assembly accuracy is verified directly on the self-consistent VASP pseudo-wavefunctions $C_\sigma(\mathbf{k})$ by evaluating the generalized eigenproblem residual $\max_n\norm{(\widetilde{H}_\sigma-E_{n\mathbf{k}\sigma}O)C_{n\sigma}}_2$ and Rayleigh-quotient deviation $\max_n|C_{n\sigma}\tr(\widetilde{H}_\sigma-E_{n\mathbf{k}\sigma}O)C_{n\sigma}|$ over occupied and low-lying conduction bands. In collinear ferromagnetic $\mathrm{Fe}$, $\mathrm{Ni}$, and $\mathrm{Co}$, maximum eigenproblem residuals are only $0.164$, $0.089$, and $0.334\text{ meV}$, respectively, confirming accurate assembly of the local potential and spin-resolved PAW one-center strength matrix $D_\sigma$.

For large unit cells sampled on coarse $\mathbf{k}$ meshes, we also tested a split-mesh scheme on $3^3$-mesh monoclinic $\beta\text{-}\mathrm{Ga_2O_3}$ that separates the fine-mesh direct operator $H_{A,\mathrm{direct}}$ from the smooth, low-rank coarse-mesh spectral-filtering correction $\Delta H_A = H_{A,\mathrm{eff}} - H_{A,\mathrm{direct}}$. Because only $\Delta H_A$ is Fourier interpolated on the coarse $3^3$ grid while exact short-range kinetic, local, and nonlocal components are preserved, the off-grid four-band RMS drops dramatically from $2138.22\text{ meV}$ (whole-matrix interpolation of $H_{A,\mathrm{eff}}$) to $9.43\text{ meV}$.

\subsection{Alternative eigenvalue calibration and symmetry reduction}
As an alternative benchmark to the spectral-filtered Ritz projection [Eqs.~\eqref{eq:filtered_G}--\eqref{eq:filtered_h} of the main text], eigenvalues of the direct-operator generalized eigenproblem $H_P X = S_P X \Lambda$ ($X\tr S_P X = I$) can be calibrated directly on the source mesh. Setting $U=S_P^{1/2}X$ and $M_0=X\tr A$, we apply Gaussian energy filtering to the leading block $N=\min(N_{\mathrm{AO}},N_b)$ followed by polar decomposition to prevent band-index swapping across crossings:
\begin{equation}
M_{ij} = M_{0,ij}\exp\!\left[-\frac{(\lambda_i-\bar{\lambda}_j)^2}{2w^2}\right] = L\Sigma V\tr,\qquad W = L V\tr,
\label{eq:supp_polar_align}
\end{equation}
where $\bar{\lambda}_j$ is the mean Ritz energy within the corresponding degenerate multiplet and $w=1\text{ eV}$. Defining the fully calibrated plateau size $p=\min(N,\lceil N_e/2\rceil+8)$, linear ramp $t_j = \min\{1,(N-j)/\max(1,N-p)\}$, and calibrated levels $\tilde{\lambda}_j = \lambda_j + t_j(E_{j\mathbf{k}} - \lambda_j)$, pulling back gives $H_A(\mathbf{k})=S_A(\mathbf{k})^{1/2}\widetilde{U}\,\mathrm{diag}(\tilde{\lambda})\,\widetilde{U}\tr S_A(\mathbf{k})^{1/2}$ with $\widetilde{U}=U\,\mathrm{diag}(W,I)$. On $21\times 21\times 1$ AB bilayer graphene across 32 off-grid $\mathbf{k}$ points, this calibration route achieves a four-band MAE of $14.96\text{ meV}$, whereas our spectral-filtered Ritz method further lowers it to $4.79\text{ meV}$. When only a finite-band spectral sum $H_{\mathrm{spec}}=A[E-\eta I]A\tr+\eta S_P$ is available, factoring $S_P=LL\tr$ and defining $Q=\mathrm{polar}(L^{-1}A_{:m})$, $T_\perp=I-QQ\tr$, and $h_{\mathrm{spec}}=L^{-1}H_{\mathrm{spec}} L^{-\dagger}$ yields the controlled orthogonal-complement operator $H_{\mathrm{ctl}}(\mathbf{k})=L[Q E_{1:m}Q\tr + T_\perp h_{\mathrm{spec}} T_\perp]L\tr$. For spinless time-reversal-invariant systems in a real spherical-harmonic basis, $M(\mathbf{k})=M(-\mathbf{k})^*$, so taking $\mathrm{Re}\,H_A(\mathbf{R})$ cleanly eliminates residual imaginary numerical noise.

\subsection{Spinor alignment under arbitrary quantization axes}\label{sec:supp_spinor}
In the window-first two-step polar orthonormalization [Eq.~\eqref{eq:protected_polar} of the main text], each source grid point $\mathbf{k}$ retains $m_{\mathbf{k}}=t+2$ bands with local degeneracy completion within a $10^{-4}\text{ eV}$ tolerance and enforces a minimum normalized singular value $\sigma_{\min}>10^{-6}$. This pointwise construction avoids transitive degeneracy inflation across $\mathbf{k}$ points while preserving machine-precision covariance under arbitrary unitary rotations within the target and buffer subspaces.

For two-component noncollinear spinor and SOC calculations performed along a nontrivial VASP spin quantization axis $\hat{\mathbf{s}}=(x,y,z)$, with azimuth $\alpha=\mathrm{atan2}(y,x)$ and polar angle $\beta=\mathrm{atan2}(\sqrt{x^2+y^2},z)$, left-multiplying the projection matrix by the $SU(2)$ spinor rotation aligns all spinors uniformly to the global Cartesian axis $(0,0,1)$:
\begin{equation}
\mathcal{U}_{\mathrm{spin}} = U(\alpha,\beta)\otimes I_{N_{\mathrm{AO}}},\qquad
U(\alpha,\beta) = e^{-i\alpha\sigma_z/2}\,e^{-i\beta\sigma_y/2}.
\label{eq:supp_su2_saxis}
\end{equation}
Rotating the quantization axis on diamond $\mathrm{Si}$ and noncollinear frustrated triangular $\mathrm{Cr_3}$ and pulling back via Eq.~\eqref{eq:supp_su2_saxis} yields maximum matrix-element differences in $H_A(\mathbf{k})$ below $1.14\times 10^{-7}$ and $9.64\times 10^{-8}\text{ eV}$, respectively, with strictly zero difference in $S_s(\mathbf{k})$.

\section{Supplemental benchmarks and ablation studies}\label{sec:supp_benchmarks}

All supplemental benchmarks in this section are performed on PW-PAW VASP calculations.

\subsection{Plane-wave completeness and external-code comparisons}
To verify numerically the completeness closure limit $S_P = A A\tr$ for $A(\mathbf{k})=\Phi\tr O C$ and $S_P(\mathbf{k})=\Phi\tr O\Phi$ [Eq.~\eqref{eq:paw_metric} of the main text], we diagonalized the entire finite plane-wave space ($N_b=459$ bands at a $300\text{ eV}$ cutoff) at the $\Gamma$ point of VASP diamond $\mathrm{Si}$. As the retained band count increases through $32$, $96$, $256$, and $459$, the closure residual $\norm{S_P - A A\tr}_F / \norm{S_P}_F$ decays monotonically through $0.00859$, $0.00482$, $0.000234$, and $3.80\times 10^{-7}$, while the norm error of the $\eta=0$ spectral-sum Hamiltonian relative to the full-spectrum limit drops from $9.04\%$ (32 bands) and $6.79\%$ (96 bands) to $0.65\%$ (256 bands) and vanishes identically at 459 bands. This controlled experiment confirms that finite-band spectral-sum error stems solely from missing high-energy plane-wave states, motivating the direct-operator complement floor $K_0$ in Eq.~\eqref{eq:filtered_K} of the main text.

Table~\ref{tab:supp_external} details the full same-source VASP comparisons against LOBSTER~\cite{maintz2016,nelson2020}, PAOFLOW~\cite{agapito2016,nardelli2018}, and HPRO~\cite{gong2024hpro} on perturbed Si ($\mathrm{Si5}$), AB-stacked bilayer graphene, and 2H bilayer $\mathrm{MoS_2}$.

\begin{table}[ht]
\centering
\caption{Detailed off-grid error comparisons (in meV) between this work and LOBSTER, PAOFLOW, and HPRO on identical VASP calculations and off-grid test meshes.}
\label{tab:supp_external}
\resizebox{\textwidth}{!}{%
\begin{tabular}{llccc}
\toprule
System (source mesh) & Reconstruction method & Off-grid band error (RMS or MAE) & Max error $\Delta_{\max}$ & Wf.\ residual RMS \\
\midrule
Perturbed Si $\mathrm{Si5}$ ($8^3$) & LOBSTER & $24.61$ (RMS) & $50.78$ & --- \\
Perturbed Si $\mathrm{Si5}$ ($8^3$) & PAOFLOW & $40.84$ (RMS) & $165.26$ & --- \\
Perturbed Si $\mathrm{Si5}$ ($8^3$) & This work & $7.24$ (RMS) & $11.62$ & $258.5$ \\
\midrule
AB bilayer graphene ($21^2\times 1$) & LOBSTER & $129.66$ (MAE) & $2950.63$ & --- \\
AB bilayer graphene ($21^2\times 1$) & PAOFLOW & $135.61$ (MAE) & $1855.46$ & --- \\
AB bilayer graphene ($21^2\times 1$) & This work & $6.16$ (MAE) & $82.57$ & $148.7$ \\
\midrule
2H bilayer $\mathrm{MoS_2}$ ($9^2\times 1$) & HPRO (unfiltered direct operator) & $89.61$ (RMS) & $138.58$ & $1094.0$ \\
2H bilayer $\mathrm{MoS_2}$ ($9^2\times 1$) & HPRO + our spectral filter & $3.54$ (RMS) & $14.98$ & $121.0$ \\
2H bilayer $\mathrm{MoS_2}$ ($9^2\times 1$) & This work (PAW spectral-filtered) & $2.76$ (RMS) & $8.69$ & $221.0$ \\
\bottomrule
\end{tabular}}
\end{table}

\subsection{Magnetic, correlated, spin-orbit, and hybrid-functional benchmarks}

Tables~\ref{tab:supp_magnetic_dftu} and~\ref{tab:supp_hse_comparison} report the complete numerical metrics corresponding to Figs.~\ref{fig:multifunctional}(b) and~\ref{fig:multifunctional}(c) of the main text, including wavefunction action residuals for magnetic and correlated systems as well as single-mesh and maximum single-state errors for HSE06 hybrid-functional calculations.

\begin{table}[ht]
\centering
\caption{Off-grid reconstruction errors (in meV) for representative ferromagnetic metals, strongly correlated DFT+$U$ oxides, and a noncollinear SOC magnet from VASP PW-PAW calculations, corresponding to Fig.~\ref{fig:multifunctional}(b) of the main text.}
\label{tab:supp_magnetic_dftu}
\resizebox{0.88\textwidth}{!}{%
\begin{tabular}{llccc}
\toprule
System & Physical regime & Band RMS & Max error $\Delta_{\max}$ & Wf.\ residual RMS \\
\midrule
bcc $\mathrm{Fe}$ & Collinear FM metal & $4.12$ & $12.64$ & $54.7$ \\
fcc $\mathrm{Co}$ & Collinear FM metal & $4.49$ & $13.82$ & $52.3$ \\
fcc $\mathrm{Ni}$ & Collinear FM metal & $4.67$ & $15.11$ & $53.9$ \\
$\mathrm{ZnO}$ & $3d$ DFT+$U$ & $1.90$ & $4.12$ & $16.8$ \\
$\mathrm{NiO}$ & AFM $3d$ DFT+$U$ & $0.96$ & $1.93$ & $15.0$ \\
$\mathrm{MnO}$ & FM $3d$ DFT+$U$ & $1.30$ & $3.01$ & $17.9$ \\
$\mathrm{CeO_2}$ & $4f$ DFT+$U$ & $3.00$ & $8.35$ & $90.5$ \\
Triangular $\mathrm{Cr_3}$ & Noncollinear +$U$+SOC & $4.79$ & $11.06$ & $157.8$ \\
\bottomrule
\end{tabular}}
\end{table}

\begin{table}[ht]
\centering
\caption{Off-grid band RMS and maximum errors $\Delta_{\mathrm{RMS}}/\Delta_{\max}$ (in meV) for VASP HSE06 hybrid-functional calculations across reconstruction methods, corresponding to Fig.~\ref{fig:multifunctional}(c) of the main text.}
\label{tab:supp_hse_comparison}
\resizebox{0.92\textwidth}{!}{%
\begin{tabular}{lcccc}
\toprule
System (mesh) & LOBSTER & PAOFLOW & Single-mesh & Dual-mesh (ours) \\
\midrule
$\mathrm{GaAs}$ ($4^3$) & $107.0 / 365.4$ & $73.6 / 205.8$ & $13.7 / 52.9$ & $2.74 / 11.87$ \\
$\mathrm{GaAs}$ ($5^3$) & $48.5 / 197.8$ & $22.9 / 75.7$ & $5.4 / 30.4$ & $1.71 / 6.68$ \\
$\mathrm{Si}$ ($4^3$) & $142.8 / 422.5$ & $78.9 / 222.7$ & $49.7 / 235.1$ & $9.97 / 38.23$ \\
$\mathrm{Si}$ ($5^3$) & $325.6 / 2090.6$ & $54.0 / 351.8$ & $23.8 / 119.3$ & $8.03 / 28.62$ \\
$\mathrm{MgO}$ ($4^3$) & $2619.5 / 12249$ & $61.9 / 172.3$ & Rank-def. & $3.05 / 12.14$ \\
$\mathrm{MgO}$ ($5^3$) & $1141.1 / 6538.2$ & $147.4 / 985.1$ & Rank-def. & $2.06 / 8.22$ \\
\bottomrule
\end{tabular}}
\end{table}

\subsection{Basis conditioning and amorphous supercell projectability}

\begin{figure}[ht]
\centering
\includegraphics[width=0.92\textwidth]{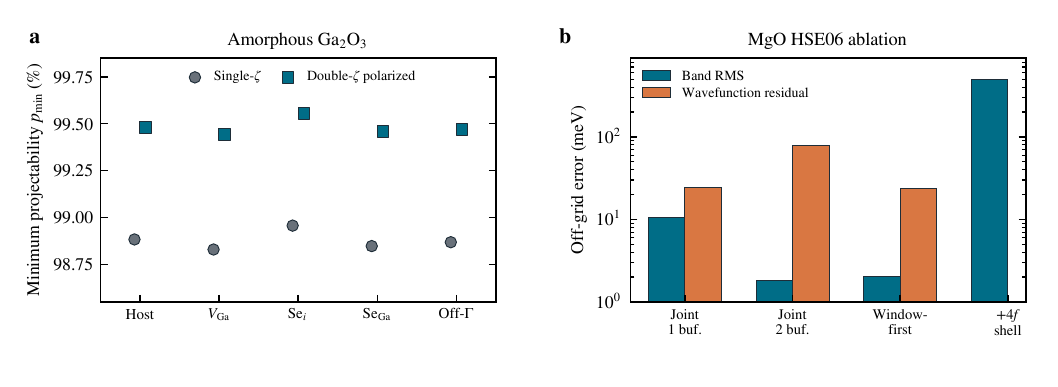}
\caption{(a) Minimum band-edge projectability $p_{\min}$ across 160-atom spin-polarized amorphous $\mathrm{Ga_2O_3}$ host and defect supercells for single-$\zeta$ and double-$\zeta$ polarized bases. (b) HSE06 off-grid band errors and wavefunction residuals in $\mathrm{MgO}$ across polar-orthonormalization schemes and $+4f$ basis augmentation.}
\label{fig:supp_amorphous_mgo}
\end{figure}

\begin{table}[ht]
\centering
\caption{Off-grid band errors, wavefunction residuals (in meV), and maximum geometric-overlap condition numbers $\kappa_{\max}(S_A)$ for wide-gap $\mathrm{MgO}$ ($10^3$ PBE anchoring $5^3$ HSE06) across polar-orthonormalization schemes and basis sizes.}
\label{tab:supp_mgo_ablation}
\resizebox{\textwidth}{!}{%
\begin{tabular}{lcccc}
\toprule
Orthonormalization and basis setup & Band RMS & Max error $\Delta_{\max}$ & Wf.\ residual RMS & $\kappa_{\max}(S_A)$ \\
\midrule
Joint polar (1 buffer band, 26 AO) & $10.65$ & $102.87$ & $24.41$ & $374.9$ \\
Joint polar (2 buffer bands, 26 AO) & $1.84$ & $7.60$ & $79.30$ & $374.9$ \\
Window-first polar (this work, 26 AO) & $2.06$ & $8.22$ & $23.77$ & $374.9$ \\
Overcomplete $+4f$ shell (joint polar, 40 AO) & $490.37$ & $3268.55$ & --- & $1999.0$ \\
\bottomrule
\end{tabular}}
\end{table}

Figure~\ref{fig:supp_amorphous_mgo}(b) and Table~\ref{tab:supp_mgo_ablation} present ablation tests for wide-gap $\mathrm{MgO}$ under VASP HSE06. Appending an extra $4f$ shell on both $\mathrm{Mg}$ and $\mathrm{O}$ expands the unit-cell dimension from 26 to 40 orbitals, degrading the maximum off-grid overlap condition number $\kappa_{\max}(S_A)$ from $374.9$ to $1999.0$; amplification of interpolation noise through $S_A^{-1/2}$ inflates the off-grid band RMS to $490.37\text{ meV}$. Maintaining the well-conditioned 26-orbital compact basis with window-first two-step polar orthonormalization [Eq.~\eqref{eq:protected_polar} of the main text] simultaneously achieves an off-grid band RMS of $2.06\text{ meV}$ and a wavefunction residual of $23.77\text{ meV}$. Across the four 160-atom collinear spin-polarized amorphous $\mathrm{Ga_2O_3}$ host and defect supercells ($\sim 2400$ orbitals per spin) shown in Fig.~\ref{fig:supp_amorphous_mgo}(a), our double-$\zeta$ polarized basis maintains minimum band-edge four-state projectabilities of $99.44\%\text{--}99.56\%$ in both spin channels.

\subsection{Energy-reference gauge and machine-learning benchmarks}

Table~\ref{tab:supp_si_transfer} lists the coupled-strain Si energy-reference gauge data plotted in Fig.~\ref{fig:transfer_gauge}(d) of the main text, and Table~\ref{tab:supp_mlh_full} provides the full 384-structure end-to-end MLH learning-curve and HSE06 complement-intrusion diagnostics corresponding to Figs.~\ref{fig:transfer_gauge}(a)--\ref{fig:transfer_gauge}(c) of the main text.

\begin{table}[ht]
\centering
\caption{Direct reconstruction errors, cross-configuration real-space matrix-element prediction errors, and off-grid band prediction RMS errors (in meV) on unseen coupled-strain Si configurations under different energy-reference gauges.}
\label{tab:supp_si_transfer}
\resizebox{\textwidth}{!}{%
\begin{tabular}{lccccc}
\toprule
Test config. & Direct AO RMS / Max & Matrix $\mathrm{RMS}(\Delta H_A)$ & Unperturbed ref.\ RMS & Floating $H_A-E_F S_A$ RMS & Fixed VASP $H_A$ RMS \\
\midrule
Config.\ 1 ($\mathrm{Si5}$) & $7.24 / 11.62$ & $1.39$ ($0.63\%$) & $178.09$ & $108.72$ & $10.17$ \\
Config.\ 2 ($\mathrm{Si6}$) & $8.91 / 27.42$ & $0.75$ ($0.34\%$) & $119.41$ & $27.07$ & $18.36$ \\
Config.\ 3 ($\mathrm{Si7}$) & $9.11 / 24.23$ & $1.33$ ($0.60\%$) & $158.55$ & $111.34$ & $20.76$ \\
Config.\ 4 ($\mathrm{Si8}$) & $9.01 / 20.16$ & $0.95$ ($0.43\%$) & --- & --- & $9.13$ \\
Config.\ 5 ($\mathrm{Si9}$) & $8.66 / 25.96$ & $0.71$ ($0.32\%$) & --- & --- & $11.39$ \\
\bottomrule
\end{tabular}}
\end{table}

\begin{table}[ht]
\centering
\caption{End-to-end MLH off-grid band RMS errors (in meV), matrix-element MAEs, and $W$-point complement-intrusion diagnostics across the three 128-structure perturbed VASP datasets.}
\label{tab:supp_mlh_full}
\resizebox{\textwidth}{!}{%
\begin{tabular}{llccccc}
\toprule
Dataset and label & Model / training size & In-domain RMS & Extrap.\ RMS & Matrix MAE & $W$ band-5 RMS & $W$ intrusions \\
\midrule
$\mathrm{MoS_2}$ (PBE) & Direct AO label & $3.31$ & $3.25$ & --- & --- & --- \\
$\mathrm{MoS_2}$ (PBE) & Equilibrium $H_{A,0}$ & $92.12$ & $148.83$ & $3.39$ & --- & --- \\
$\mathrm{MoS_2}$ (PBE) & 6D linear response & $5.72$ & $9.52$ & $0.58$ & --- & --- \\
$\mathrm{MoS_2}$ (PBE) & HamGNN ($N_{\mathrm{train}}=18 / 40 / 80$) & $4.86 / 3.80 / 3.29$ & $7.95 / 5.70 / 3.72$ & $0.35$ & --- & --- \\
\midrule
$\mathrm{CsPbI_3}$ (PBE) & Direct AO label & $2.93$ & $3.15$ & --- & --- & --- \\
$\mathrm{CsPbI_3}$ (PBE) & Equilibrium $H_{A,0}$ & $110.68$ & $153.13$ & $0.98$ & --- & --- \\
$\mathrm{CsPbI_3}$ (PBE) & 6D linear response & $3.79$ & $5.29$ & $0.22$ & --- & --- \\
$\mathrm{CsPbI_3}$ (PBE) & HamGNN ($N_{\mathrm{train}}=18 / 40 / 80$) & $3.07 / 2.90 / 2.93$ & $3.75 / 3.58 / 3.43$ & $0.12$ & --- & --- \\
\midrule
$\mathrm{Si}$ (HSE06, conventional) & HamGNN ($N_{\mathrm{train}}=20 / 40 / 80$) & $76.9 / 60.6 / 33.1$ & $1392.0 / 410.6 / 317.3$ & $0.75$ & $1416.9$ & $11/24$ \\
$\mathrm{Si}$ (HSE06, no locality) & HamGNN ($N_{\mathrm{train}}=20 / 40 / 80$) & $183.3 / 158.2 / 157.7$ & $214.5 / 190.0 / 202.8$ & $0.26$ & $695.5$ & $2/24$ \\
$\mathrm{Si}$ (HSE06, this work) & Direct AO label & $13.47$ & $11.98$ & --- & $10.8$ & $0/24$ \\
$\mathrm{Si}$ (HSE06, this work) & HamGNN ($N_{\mathrm{train}}=20 / 40 / 80$) & $12.88 / 12.71 / 12.76$ & $12.58 / 12.51 / 13.17$ & $0.16$ & $12.3$ & $0/24$ \\
\bottomrule
\end{tabular}}
\end{table}

}

\end{document}